\documentclass[article,12pt,oneside,a4paper,english,brazil,sumario=abnt-6027-2012]{abntex2}

\usepackage{graphicx,url}
\usepackage{amsmath,amssymb}
\usepackage{verbatim}
\usepackage{siunitx}

\usepackage{endnotes}
\let\footnote=\endnote

\usepackage{lmodern}			
\usepackage[T1]{fontenc}		
\usepackage[utf8]{inputenc}		
\usepackage{indentfirst}		
\usepackage{nomencl} 			
\usepackage{color}				
\usepackage{graphicx}			
\usepackage{microtype} 

\usepackage{tabularx}

\usepackage[alf,bibjustif]{abntex2cite} 

\usepackage{fancyhdr}
\fancypagestyle{alim}{\fancyhf{}%
  \fancyhfoffset{0cm}
  \fancyhead[L]{\color{red}\scalebox{0.75}{\parbox{9cm}{\vspace{2cm}\tiny This postprint has been reformatted; the published article may be found at \url{https://www.revistas.ufg.br/musica/article/view/53570}}}}
  \fancyhead[R]{%
    \scalebox{0.6}[0.8]{\textsf{%
        SINCLAIR, S.\ Sounderfeit: Cloning
        a Physical Model using a Conditional Adversarial Autoencoder}}\\
    \scalebox{0.6}[0.8]{\textsf{%
        Revista M\'usica Hodie, Goi\^ania, V.18 - n.1, 2018, p.\ 44--60}}}
  \fancyfoot[C]{\sffamily\Large\begin{tabular}{c}\hline\thepage\\\hline\end{tabular}}
  \fancyfoot[L]{\raisebox{1.5cm}{\parbox{0.5\textwidth}{\bigskip\scalebox{0.6}[0.7]{\textsf{%
        Revista M\'usica Hodie, Goi\^ania - V.18, 165p., n.1, 2018}}}}}
  \fancyfoot[R]{\raisebox{1.5cm}{\parbox{0.5\textwidth}{\flushright\scalebox{0.6}[0.7]{\textsf{%
        Recibido em: 27/11/2017 - Aprovado em: 05/03/2018}}}}}
  \setlength{\headheight}{28pt}
}

\makeatletter
\def\thickhline{%
  \noalign{\ifnum0=`}\fi\hrule \@height \thickarrayrulewidth \futurelet
   \reserved@a\@xthickhline}
\def\@xthickhline{\ifx\reserved@a\thickhline
               \vskip\doublerulesep
               \vskip-\thickarrayrulewidth
             \fi
      \ifnum0=`{\fi}}
\makeatother

\newlength{\thickarrayrulewidth}
\fancypagestyle{plain}{\fancyhf{}%
  \fancyhfoffset{0cm}
  \fancyhead[R]{%
    \scalebox{0.6}[0.8]{\textsf{%
        SINCLAIR, S.\ Sounderfeit: Cloning
        a Physical Model using a Conditional Adversarial Autoencoder}}\\
    \scalebox{0.6}[0.8]{\textsf{%
        Revista M\'usica Hodie, Goi\^ania, V.18 - n.1, 2018, p.\ 44--60}}}
  \fancyfoot[C]{\sffamily\Large\begin{tabular}{c}\thickhline\thepage\\\thickhline\end{tabular}}
}

\fancypagestyle{last}{\fancyhf{}%
  \fancyhfoffset{0cm}
  \fancyhead[R]{%
    \scalebox{0.6}[0.8]{\textsf{%
        SINCLAIR, S.\ Sounderfeit: Cloning
        a Physical Model using a Conditional Adversarial Autoencoder}}\\
    \scalebox{0.6}[0.8]{\textsf{%
        Revista M\'usica Hodie, Goi\^ania, V.18 - n.1, 2018, p.\ 44--60}}}
  \fancyfoot[C]{%
    {\vspace{-1.2cm}\tiny
      \color{red}(This postprint has been reformatted from the published article)} \\~\\
    \sffamily\Large\begin{tabular}{c}\thickhline\thepage\\\thickhline\end{tabular}}
}

\graphicspath{{../results-ey0.1-lr0.1/}}

\titulo{\vspace{-1cm}\large\textbf{Sounderfeit: Cloning a Physical Model using a Conditional
    Adversarial Autoencoder}}

\autor{Stephen Sinclair (Inria Chile, Santiago, Chile) \\ stephen.sinclair@inria.cl}

\preauthor{\begin{flushright}}
\postauthor{\end{flushright}}

\data{}

\makeindex

\setlrmarginsandblock{2.5cm}{2.5cm}{*}
\setulmarginsandblock{2.5cm}{3.5cm}{*}
\checkandfixthelayout

\SingleSpacing

\captionnamefont{\sffamily\smaller}
\captiontitlefont{\sffamily\smaller\hangindent=0cm}

\makeatletter
\renewcommand{\@seccntformat}[1]{\csname the#1\endcsname. }
\makeatother

\hypersetup{hidelinks}

\begin{document}
\setcounter{page}{44}
\maketitle
\pagestyle{fancy}
\thispagestyle{alim}

\setulmarginsandblock{2.5cm}{2.5cm}{*}
\checkandfixthelayout

\selectlanguage{english}

\frenchspacing

\renewcommand\resumoname{\vspace{-5em}}
\renewcommand\bibname{References}

\vspace{-1cm}

\begin{resumoumacoluna}
\textbf{Abstract:}
An adversarial autoencoder conditioned on known parameters of a
physical modeling bowed string synthesizer is evaluated for use in
parameter estimation and resynthesis tasks. Latent dimensions are
provided to capture variance not explained by the conditional
parameters. Results are compared with and without the adversarial
training, and a system capable of ``copying'' a given parameter-signal
bidirectional relationship is examined.  A real-time synthesis system
built on a generative, conditioned and regularized neural network is
presented, allowing to construct engaging sound synthesizers based
purely on recorded data.


\bigskip\noindent
Sounderfeit: Clonagem de um modelo físico com auto-encoders
advers\'arios condicionais

\textbf{Resumo:}
Um autocodificador adversarial condicionado a par\^ametros conhecidos
de um sintetizador modelado f\'isico de cordas arqueadas \'e avaliado
por sua utiliza\c{c}\~ao na estima\c{c}\~ao de par\^ametros e nas
tarefas de res\'intese. O dimensões latentes s\~ao fornecidas para
capturar a vari\^ancia n\~ao explicado pelos par\^ametros
condicionais. Os resultados s\~ao comparados com e sem o treinamento
advers\'ario, e um sistema capaz de ``copiar'' uma determinada
rela\c{c}\~ao bidirecional entre par\^ametros e sinal é examinado. Um
sistema de s\'intese em tempo real constru\'ido em uma rede neuronal
generativa, condicionada e regularizada \'e apresentada, permitindo
criar sintetizadores de som interessantes baseados puramente em dados
gravados.

\textbf{Palavras-chave:} Modelagem f\'isica,
s\'intese de som, autocoder, par\^ametros latentes
\end{resumoumacoluna}

\textual

\section{Introduction}

This paper explores the use of an autoencoder to mimic the
bidirectional parameter-data relationship of an audio synthesizer,
effectively ``cloning'' its operation while regularizing the
parameter space for interactive control.
The \emph{autoencoder}\footnote{Audio examples at:
  \url{https://emac.ufg.br/up/269/o/Sinclair_soundexample.mp3}.} is an
artificial neural network (ANN) configuration in which the network
weights are trained to minimize the difference between input and
output, learning the identity function.
When forced through a bottleneck layer of few parameters, the network
is made to represent the data with a low-dimensional ``code,''
which we call the latent parameters.

Recently adversarial configurations have been proposed as a method of
regularizing this latent parameter space in order to match it to
a given distribution \cite{makhzani2015adversarial}.
The advantages are two-fold: to ensure the available range is
uniformly covered, making it a useful interpolation space; and to
maximally reduce correlation between parameters, encouraging them to
represent orthogonal aspects of the variance.
For example, in a face-generator model, this could translate to
parameters for hair style and the presence of glasses
\cite{radford2015unsupervised}.
Meanwhile, it has also been shown that a generative network can be
conditioned on known parameters \cite{Mirza2014}, to
make it possible to control the output, for example, to generate a
known digit class when trained on MNIST digits.

In this work, these two concepts are combined to explore whether an
adversarial autoencoder can be conditioned on known parameters for use
in both parameter estimation and synthesis tasks for audio.
In essence, we seek to have the network simultaneously learn to mimic
the transfer function from parameters to data of a periodic signal, as
well as from data to parameters.
Latent dimensions are provided to the network to capture variance not
explained by the conditional parameters; in audio, they may represent
internal state, stochastic sources of variance, or unrepresented
parameters e.g.\ low-frequency oscillators.
The idea of using adversarial training to regularize the distribution
of the latent space is to find a configuration such that the
parameters are made to lie in a predictable range and uniformly fill
the space, in order to provide a system suitable for live interaction.

For the principal test case herein, we train the autoencoder on
waveform periods from a physical modeling synthesizer based on a model
of the interaction between a string and a bow. The goal is to produce
a black-box parameter estimator and synthesizer that both ``listens
to'' (estimates physical parameters) of an incoming sound and
reproduces it, with a parameter space optionally informed by the
original parameters. Application of the architecture described here is
of course not limited to physical models, but may be applied to any
periodic sound source; a physical model was chosen for its ability to
produce fairly complex signals from a simple parameter mapping, and
the periodic requirement comes mainly from needing a constant size for
the input and output network layers. Results are visualized and some
informal qualitative evaluations are discussed. The autoencoder was
able to reproduce the steady state of the synthesizer with and without
regularization, although reproduction error increased, expectedly, in
the presence of regularization. Some parameter estimation problems
were identified with the dataset and sampling method used, and we
conclude with some lessons learned in the art of ``synth cloning''. A
real-time system, Sounderfeit, built on a generative neural network is
presented, allowing to construct engaging sound synthesizers based
purely on recorded data and optional prior knowledge of parameters.

\section{Previous work}

Previous publication of this work \cite{Sinclair2017sbcm} did not fully
compare the results with related literature in the audio domain, and
therefore in this extended version we include a more thorough overview
of related work in this section.  Indeed this work combines two ideas
that have been previously investigated, that of parameter estimation,
and that of ANN-based audio synthesis.

Note that in the following we skip mention of several works that make
use of similar ANN approaches for classifying sounds; in
fact quite a lot of this work is available in the music information
retrieval literature, and thus we restrict the discussion to papers
that specifically discuss parameter estimation and audio synthesis.
Parameter estimation for physical modeling is a well-researched topic,
however it typically leverages known relations between observable
aspects of the signal and physically-relevant parameters,
c.f.~\cite{Scherrer2011}.  The application of black-box, ANN-based
models is until recently rather less common, but several works can be
found in the literature.  The intuition in such an approach is that
since a physical model represents a non-linear, stateful
transformation of the parameters, the generated signal tends to be
difficult to separate into effects originating from specific
parameters, thus an arbitrary non-linear multivariate regression based
on known data is a more pragmatic approach to constructing such an
\emph{inverse mapping}.

For example, \citeonline{Cemgil1997} investigated the application of ANN for
estimating the parameters of a plucked string model.
Also in this vein, \citeonline{Riionheimo2003} used a genetic search
strategy for finding similar parameters.  They employed a perceptual
model as their error metric in order to better measure distance
between sets of parameters as perceived by humans.  A parameter space
quantized (also according to a perceptual model) was used.
Similarly, \citeonline{Gabrielli2017param} used a multilayered
convolutional neural network (deep CNN) to determine parameters of
an organ physical model that supports up to 58 parameters (organ
stops) per key.  Short-time Fourier spectra were used as input
calculated from a dataset of 2220 samples, and the network was trained
to minimize the mean squared error on the parameter reconstruction.  A
notion of \emph{spectral irregularity} was used to judge the
similarity of resulting synthesized sounds.
\citeonline{pfalz2017toward} explored the use of a long short-term memory
recursive neural network (LSTM-RNN) to estimate the continuous control
signals from the output of a physical model.  Mean squared error of
the reconstructed parameters to the original parameters is used as the
loss.  It was successful at determining trigger times
and parameters for fairly simple plucked gestures with several types
of resonator models over a few seconds of time, but generated spurious
triggers when trained on more complex musical gestures.

Regarding generation of audio using neural networks, generally two
approaches are used: either (1) generation of pulse-coded audio one
sample at a time using a sequential model, e.g.~an autoregressive
model or a recursive neural network (RNN); or (2) generation of audio
frames in the form of spectra or spectrograms (series of spectra).
The current work takes an alternative approach (3) generation of
pulse-coded time-domain audio frames.

An example of the first technique, sample-at-a-time synthesis, is
WaveNet \cite{oord2016wavenet}, in which a multilayered CNN with
exponentially dilated receptive fields is used to model progressively
short- to long-term dependencies in the audio stream.  The ``receptive
field'' is enlarged exponentially at each layer using dilated causal
convolutions.  This configuration may be conditioned on external
variables, for example speaker identification, phoneme information,
musical style, etc.
In addition to testing this model on speech coding, it was later used
in an autoencoder configuration, dubbed NSynth, in a way quite
comparable to the current work, that is, to encode and reproduce
musical instrument tones \cite{engel2017neural}.  Their analysis goes
into depth on the qualities of reconstruction compared to a baseline
model, which itself is also a deep CNN described below, and describes
some temporal aspects of the learned latent space, as well as the
qualitative effects of latent-space interpolation.

In comparison, another example is SampleRNN \cite{Mehri2016samplernn},
which used multi-scaled deep RNNs to capture long-term dependencies as
a stacked autoregressive model, i.e.,~it encodes the conditional
probability distribution of the next sample based on previous samples
and encodings produced by other layers.  The multiple scales allow
this sample-at-a-time model to also take into account frame-level
information, and can therefore use higher levels to encode longer-term
dependencies.
It was compared to WaveNet and a standard RNN with Gaussian mixture
model in terms of reconstruction mean squared error, and with human
listening preference experiments on encodings of voice, human
non-vocal sounds, and piano music, and performed favourably.
Interestingly, rather than purely real-valued output, all of the
above-mentioned sample-at-a-time methods made use of a quantized
one-hot categorical softmax over an 8-bit $\mu$-law encoding for
estimating the real value of the audio signal.  The intuition is that
such an encoding allows to remove any prior assumptions about
distribution and simply take the most probable discrete value.

As for frame-at-a-time methods, the baseline model from
\citeonline{engel2017neural} is applicable, as it consists of a deep CNN
trained on spectrograms.  As mentioned, they used a large instrument
dataset and a large latent space of approximately 2000 dimensions to
encode both the time and frequency domains.  Indeed, this model can be
thought of more as a spectrogram-at-a-time rather than
frame-at-a-time, since the encoding takes multiple frames into
account; the latent dimensions per frame were on the order of 16 and
32.  The authors reported poor performance for encoding phase or
complex representations, and thus used only spectral magnitude as
input, and applied a phase reconstruction technique to synthesize the
final audio.

In a work similar in motivation to the current one,
\citeonline{Riera2017sbcm} used a sparse autoencoder to generate a set of
descriptors according to the latent space of the model.  A
multilayered network was used with a latent model of 8 dimensions,
trained on all frames of a single recording.  The sparse activation of
the central bottleneck layer is visualized in time and interpreted as
a ``neural score'', and can be used to reconstruct the original audio.
A comparison of the clustering in the first three principle components
is provided to compare the resulting ``timbre space'' in terms of
activations of the latent layer with those of MFCC and spectral
contrast descriptors, which are organized qualitatively differently;
the former show a distinctly less ``cloudy'' shape compared to the
latter, and instead distinct curved lines or trajectories are
apparent.  This is of course qualitatively open to interpretation, but
does imply some kind of structure imposed on the latent space that
appears to differ significantly from the use of non-learned
descriptors.
In this work, we found similar patterns in unregularized latent
spaces, e.g.\ Figures \ref{bowed-paramdist}a and
\ref{bowed-paramdist-tanh}a.

A distinguishing factor of this work compared to NSynth is that rather
than attempt to model a large set of instruments, which requires a
large model, large dataset, and large-dimensional latent space (16 or
32) with unknown meaning, we focus on representing the sound with a
comparatively small set of parameters (2 to 3) and attempt to learn a
minimal encoding based on previous knowledge of the model parameters,
adding latent dimensions only as necessary. This stems from a
different motivation, which, instead of being to determine
multi-instrument embedding spaces as in the case of NSynth, is to
better understand the inverse data-parameter relationship, as well as
to provide a small, salient set of ``knobs'' for real-time synthesis
of a single family of timbres.

\section{Datasets}

Given a network with sufficient capacity we can encode any functional
relationship, but for the experiments described herein a periodic
signal specified by a small number of parameters was sought that
nonetheless features some complexity and is related to sound
synthesis.
Thus, a physical modeling synthesizer proved a good choice.
We used the bowed string model from the \emph{STK Synthesis Toolkit in
  C++} \cite{CookS99}, which uses digital waveguide synthesis and is
controlled by 4 parameters: \emph{bow pressure}, the force of the bow
on the string; \emph{bow velocity}, the velocity of the bow across the
string; \emph{bow position}, the distance of the string-bow
intersection from the bridge; and \emph{frequency}, which controls the
length of the delay lines and filter parameters, and thus the tuning
of the instrument.

The parameters are represented in STK as values from 0 to 128, and
thus we do not worry about physical units in this paper; all
parameters were linearly scaled to a range $[-1,1]$ for input to the
neural network.
The data was similarly scaled for input, and a linear descaling of the
output is performed for the diagrams in this paper.
Additionally, the per-element mean and standard deviations across the
entire dataset were subtracted and divided respectively in order to
ensure similar variance for each discrete step of the waveform period.

To extract the data, a program was written to evaluate the bowed
string model at 48000~Hz for 1~second for each combination of \emph{bow
  position} and \emph{bow pressure} for integers 0 to 128.
The 1-second interval was used to ensure the sound reached a steady
state with a constant period size.
The \emph{bow velocity} and \emph{volume} parameters were both held at
a value of 100.
For each instance, the last two periods of oscillation were kept, and
since some parameter combinations did not give rise to stable
oscillation, recordings with an RMS output lower than $10^{-5}$ (in
normalized units) over this span were rejected, giving a total of
15731 recordings evenly distributed over the parameter range.
The frequency was selected at 476.5~Hz to count 201 samples to capture
two periods---some parameter combinations changed the tuning slightly,
but inspection by eye of 50 periods concatenated end to end showed
minimal deviation at this frequency for a wide variety of parameters.
Two periods were recorded in order to minimize the impact of any
possible reproduction artifacts at the edges of the recording during
overlap-add synthesis.
The recordings were phase-aligned using a cross-correlation analysis
with a representative random sample, then differentiated by
first-order difference, and 200 sample-to-sample differences were thus
used as the training data, normalized as stated above.
This dataset we refer to as \emph{bowed1}.

Although it may be beneficial to use a log-spectrum representation
rather than ``raw'' (pulse-coded) audio \cite{engel2017neural}, we
found that learning the time-domain oscillation cycle was no problem.
In this manner we avoided the need to perform phase reconstruction.
The use of a differentiated representation also helped to suppress
noise.
As will be discussed below, parameter estimation on new data was not
successful based on this dataset due to the lack of representation
of the synthesizer's dynamic regimes.
To resolve this, a second extended dataset, \emph{bowed2}, was created
in a similar manner, however instead of recording only the steady
state portion, the synthesizer was executed continuously while
changing the parameters randomly at random intervals.
100,000 samples uniformly covering the parameter range were captured
for \emph{bowed2}.

Finally, in order to test the idea on a completely independent albeit
simple dataset, a human voice was recorded uttering constant vowel
sounds.  The voice (the author's own voice) was held steady in
frequency for a period of 3 seconds for vowels \emph{a}, \emph{e},
\emph{i}, \emph{o}, and \emph{u}.
The beginning and end of each utterance was clipped and periods were
extracted and globally phase-aligned by aligning peaks.
The voice was recorded at 44100~Hz and had a frequency between 114 and
117~Hz, thus slightly long cycles were extracted to have exactly 400.5
samples per period, so that two periods were 801 samples, or 800
samples in the differential representation used here.  This created a
final fundamental frequency of 110~Hz in the synthesized sound.  (Due
to overlap-add, artifacts in 2 or 3 samples at the beginning and end
of a cycle are mostly surpressed.)
Integers 0 through 4 were assigned to each vowel and used as the single
conditional parameter.
This resulted in 996 two-period samples, or approximately 200 samples
per vowel.
As will be shown, since the voice was held quite steady, most periods
for the same vowel were quite similar, however a low-quality
microphone and natural vocal variation contributed to differences
between samples.
This dataset is referred to in this text as \emph{vowels}.

In all cases, reproduction consists of de-normalizing, concatenating
using an overlap-add method, and first-order integrating the final
signal. A 50\% overlap-add with a Hanning window was used, which
features a constant overlap summation thereby avoiding modulation
artifacts \cite{Smith1987parshl}. The parameters are assumed constant
during one window, and thus interpolation artifacts may begin to
appear if the parameters changed quickly relative to two cycles of the
waveform. In the ideal case, perfect reproduction of each cycle
concatenated using this technique should reproduce the steady-state
waveform of the original sound source

\section{Training and network architecture}

\subsection{Learned conditional autoencoding}

While the principle job of the autoencoder is to reproduce the input
as exactly as possible, in this work we also wish to estimate the
parameters used to generate the data.
Thus we additionally \emph{condition} part of the latent space by
adding a loss related to the parameter reconstruction.
This is somewhat different to providing conditional parameters to the
\emph{input} of the encoder
\cite{makhzani2015adversarial,Mirza2014}, but has a
similar effect.  This is to encourage the network to learn how to
recognize the known parameters and assign aspects of the variance to
them that is associated with those parameters.

Note that the presence of the latent parameters is what allows for the
fact that we do not assume that the signal is purely deterministic in
the known parameters.
For instance, in a physical signal there may be internal state
variables that are not taken into account in the initial conditions,
or acoustic characteristics such as room reverb that are not
considered a priori.
Naturally, the less deterministic the signal is in the known
parameters, the more must be left to latent parameters, and the poorer
a job we can expect the parameter reconstruction to do.
Note that if the latent parameters are able to represent the dynamic
regimes, then dynamical state changes may be represented as
trajectories in the latent space, however we did not try to
reconstruct such trajectories in this work.

\subsection{Generative adversarial regularization}

The code used in the middle layer of an autoencoder, called the
\emph{latent parameters}, which we shall refer to as $z$, when trained
to encode the data distribution $p(x)$, has conditional posterior
probability distribution $q(z|x)$.
As mentioned, it is in general useful to regularize $q(z|x)$ to match
a desired distribution.

Several methods exist for this purpose: a \emph{variational
  autoencoder} (VAE) uses the Kullback-Leibler divergence from a given
prior distribution.
Other measures of difference from a prior are possible.
The use of an adversarial configuration has been proposed
\cite{makhzani2015adversarial} to regularize $q(z)$ based on the
negative log likelihood from a discriminator on $z$.

With adversarial regularization, a discriminator is used to judge
whether a posterior distribution $q(z)$ was likely produced by the
generator and is thus sampled from $q(z|x)$, or rather sampled from an
example distribution $p(z)$, which is often set to a normal or uniform
distribution.
The discriminator is itself an ANN which outputs a 1 if $z$ consists
of a ``real'' sample of $p(z)$ or a 0 for a ``fake'' sample of
$q(z|x)$.  The training loss of the generator, which is also the
encoder of the autoencoder, maximizes the probability of fooling the
discriminator into thinking it is a real sample of $p(z)$, while the
discriminator simultaneously tries to increase its accuracy at
distinguishing samples from $p(z)$ and samples from $q(z|x)$.
Thus thus encoder eventually generates posterior $q(z|x)$ to be
similar to $p(z)$.

\begin{figure}
  \centering\includegraphics[width=7cm]{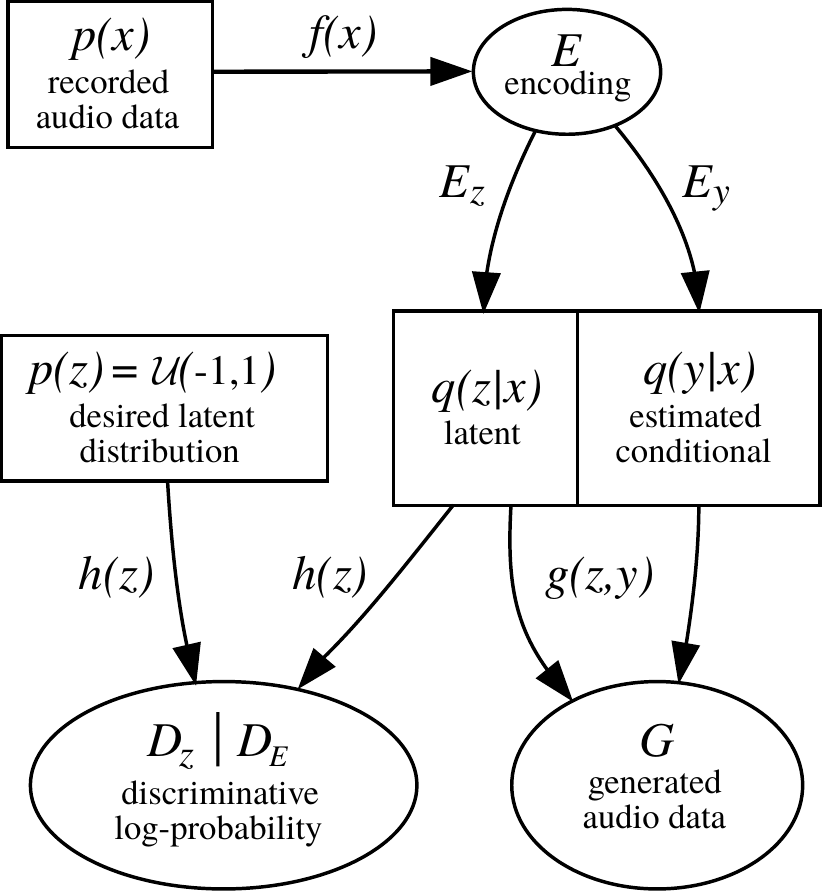}
  \caption{Variables and functions in the description of the adversarial autoencoder.
    \label{systemdiag}}
\end{figure}

\subsection{Network description}

Putting together the above concepts, the system is composed of two
neural networks and three training steps.
A visual description of the network configuration and how it relates
to the following variables and functions may be found in
Figure~\ref{systemdiag}.

First, the autoencoder network is composed of the encoder $E=f(x)$ and
the decoder/generator $G=g(z,y)$. The discriminator is designed
analogously as $D=h(z)$.
For notational convenience, we also define $G_E=g(E)=g(f(\bold{x}))$,
$D_E=h(E_z)$, and $D_z=h(\bold{z})$ where $\bold{x}=\bold{x}_1\ldots
\bold{x}_s$ are sampled from $p(x)$, $\bold{z}=\bold{z}_1\ldots
\bold{z}_s$ is sampled from $p(z)$, and $s$ is the batch size.
$E_z(x)$ and $E_y(x)$ are the first $n$ and the last $m$ dimensions of
$E \in [z_1 \cdots z_n \;\; y_1 \cdots y_m]$, respectively.
In the current work, $f(x)$ and $g(z,y)$ are simple one-hidden-layer
ANNs with one non-linearity $\zeta$ and linear outputs:
\begin{align}
  f(x)&=\zeta ( x \cdot w_1 + b_1 ) \cdot w_2 + b_2 \\
  g(z,y)&=\zeta ( [z \; y] \cdot w_3 + b_3 ) \cdot w_4 + b_4 \\
  h(z)&=\zeta ( z \cdot w_5 + b_5 ) \cdot w_6 + b_6
\end{align}
We used the rectified linear unit $\zeta(x)=\max(0,x)$ (ReLU), but we
also investigated the use of \emph{tanh} non-linearities, described in
Section~\ref{sec:results}.

The principal dataset, described below, was composed of 200-wide 1-D
vectors, and we had acceptable results using hidden layers of half
that size, so $w_1 \in \mathbb{R}^{200 \times 100}$, $w_2 \in
\mathbb{R}^{100 \times {(n+m)}}$ and $w_3,w_5 \in
\mathbb{R}^{{(n+m)}\times 100}$, $w_4,w_6 \in \mathbb{R}^{100\times
  1}$, where ${(n+m)}$, the total size of the hidden code, was 2 or 3,
depending on the experiment.  The bias vectors $b_1 \ldots b_6$ had
corresponding sizes accordingly.

\subsection{Training}

The training steps were performed in the following order for each batch:%
\footnote{ The learning rates have been changed from
  \citeonline{Sinclair2017sbcm}: stochastic gradient descent with
  learning rate 0.005 for the autoencoder and learning rate 0.05 for
  the generator and discriminator.  Due to a programming error, the
  autoencoder training step performed better with a different learning
  rate.  We later found that the results were much more robust with
  the Adam optimiser and the same learning rate value for all training
  steps.}
\begin{enumerate}
\item The Adam optimiser \cite{kingma2014adam} with a learning rate of
  0.001 was used to train the full set of autoencoder weights $w_1
  \ldots w_4$, and $b_1 \ldots b_4$, minimizing both the data $x$
  reconstruction loss and parameter $y$ reconstruction loss,
  $\mathcal{L}_\textrm{AE}$ by back-propagation.
  The weighting parameter $\lambda=0.5$ is described below.
\item Adam with learning rate 0.001 was used to train the generator
  weights and biases $w_1$, $w_2$, $b_1$, and $b_2$.
  The negative log-likelihood $\mathcal{L}_\textrm{G}$ was minimized
  by back-propagation.
\item Adam with learning rate 0.001 was used to train the discriminator
  weights and biases $w_5$, $w_6$, $b_5$, and $b_6$.
  The negative log-likeli\-hood $\mathcal{L}_\textrm{D}$ was
  minimized by back-propagation.
\end{enumerate}

where,
\begin{align}
  \mathcal{L}_\textrm{AE}&=\sum(\bold{x}-g(f(\bold{x})))^2
  + \lambda\sum(\bold{y} - g(\bold{x}))^2 \\
  \mathcal{L}_\textrm{G}&=-\sum\log(D_E) \\
  \mathcal{L}_\textrm{D}&=-\sum(\log(D_z) + \log(1-D_E)).
\end{align}

Experiments were performed using the TensorFlow framework
\cite{tensorflow2015-whitepaper}, which implemented the
differentiation and gradient descent (back-propagation) algorithms.
A small batch size of 50 was used, with each experiment evaluated
after 4,000 batches.  It was found that smaller batch sizes worked
better for the adversarial configuration.
Matrices $\bold{z}$ and $\bold{x},\bold{y}$ were sampled independently
from $\bold{Z}\sim{}p(z)=\mathcal{U}(-1,1)$ and
$(\bold{X},\bold{Y})\sim{}p(x,y)$ for each step, where
$\mathcal{U}(a,b)$ is the uniform distribution in range $[a,b]$
inclusive.

\section{Experiments}

Six conditions were tested in order to explore the role of
conditional and latent parameters.
The number of known parameters in the dataset was 2.  We tried
training the \emph{bowed1} dataset with and without an extra latent
parameter.  We label these conditions $D1_Z2_Y$ and $D0_Z2_Y$
respectively.
The third condition, $N1_Z2_Y$, was like the $D1_Z2_Y$ condition but
without adversarial regularization on $q(z|x)$.
Thus the $D$ label is to indicate the use of training on the
discriminator, while $N$ indicates \emph{No discriminator.}

To compare conditioning with the ``natural'' distribution of the data
among latent parameters and the effects of adversarial regularization
thereupon, two configurations with no conditional parameters, with and
without the discriminator, were explored, named $D2_Z0_Y$ and
$N2_Z0_Y$ respectively.

\begin{figure}[h]
  \centerline{\sffamily
    (a) \includegraphics[height=7cm]{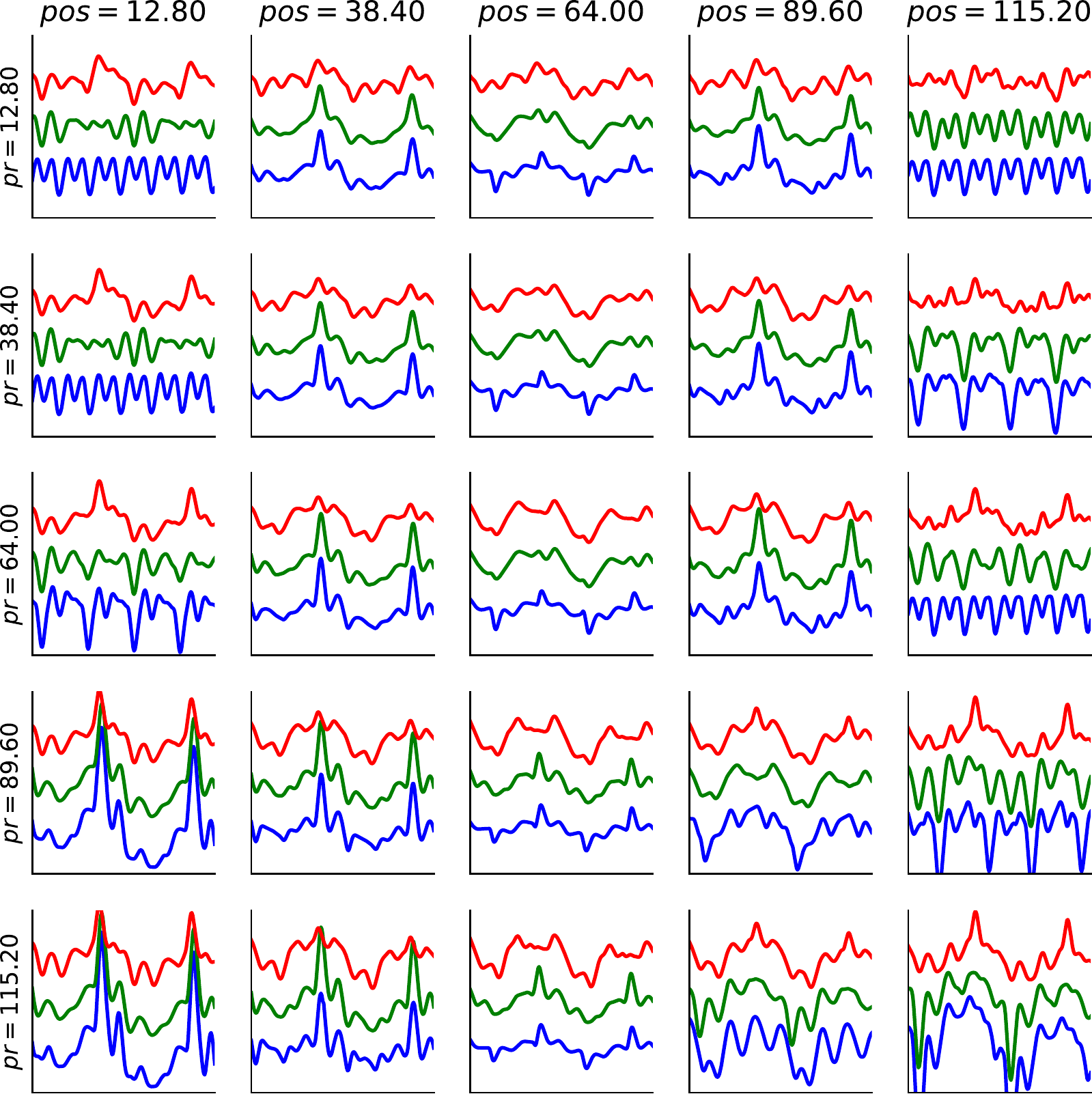} \hspace{0.2cm}
    (b) \includegraphics[height=7cm]{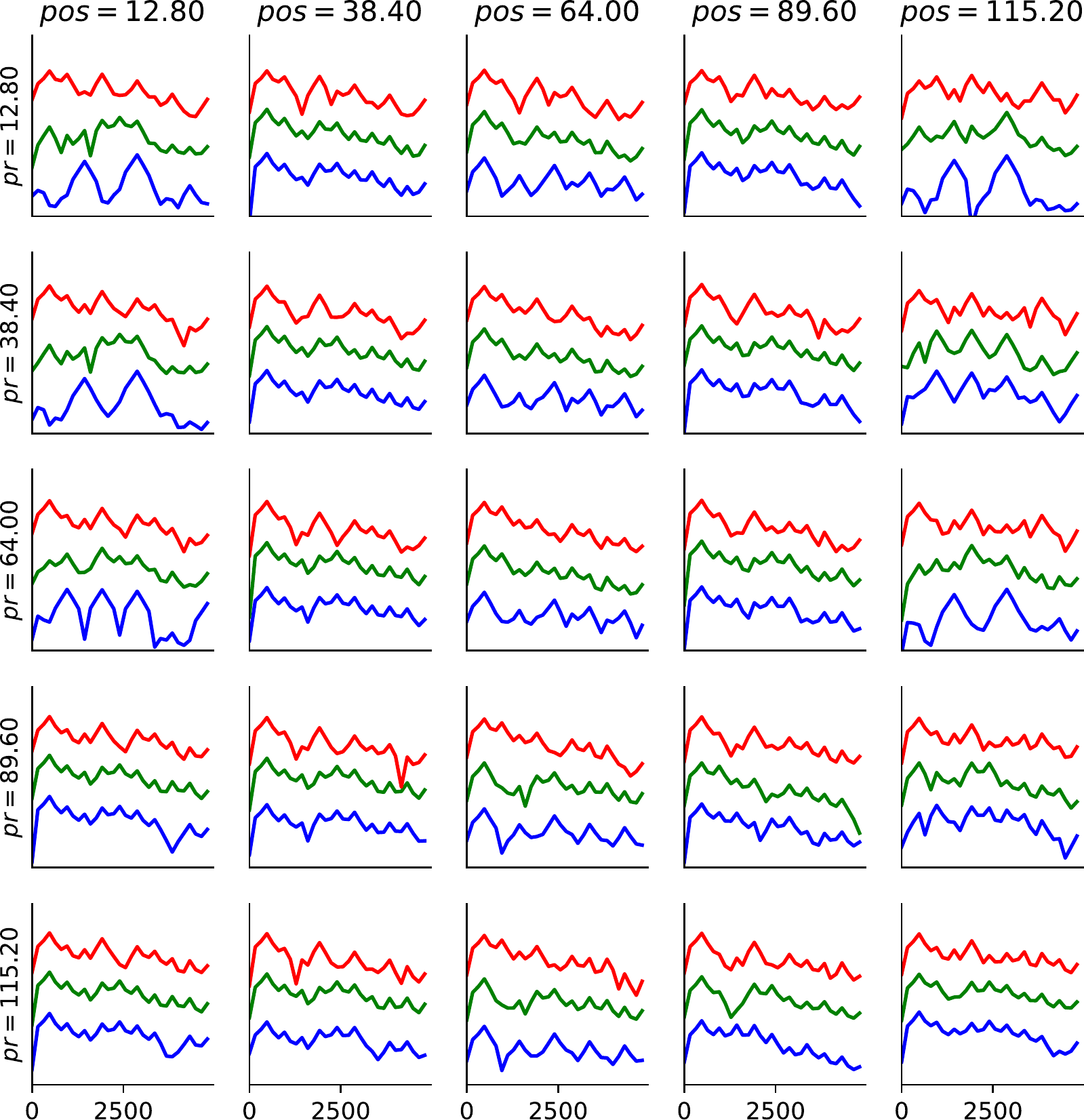}}
  \caption{ Output of $D1_Z2_Y$ as pressure $y_0$ and position $y_1$
    are changed.  Top (red) is the decoder with parameters explicitly
    specified and $z_0=0$; middle (green) is with parameters and $z_0$
    inferred by the encoder, bottom (blue) is the dataset sample with
    closest parameters.  (a) Time domain; (b) Frequency domain.  }
  \label{bowed-D1Z2Y-varypv}
\end{figure}

\begin{figure}[h]
  \centerline{\sffamily
    (a) \includegraphics[height=7cm]{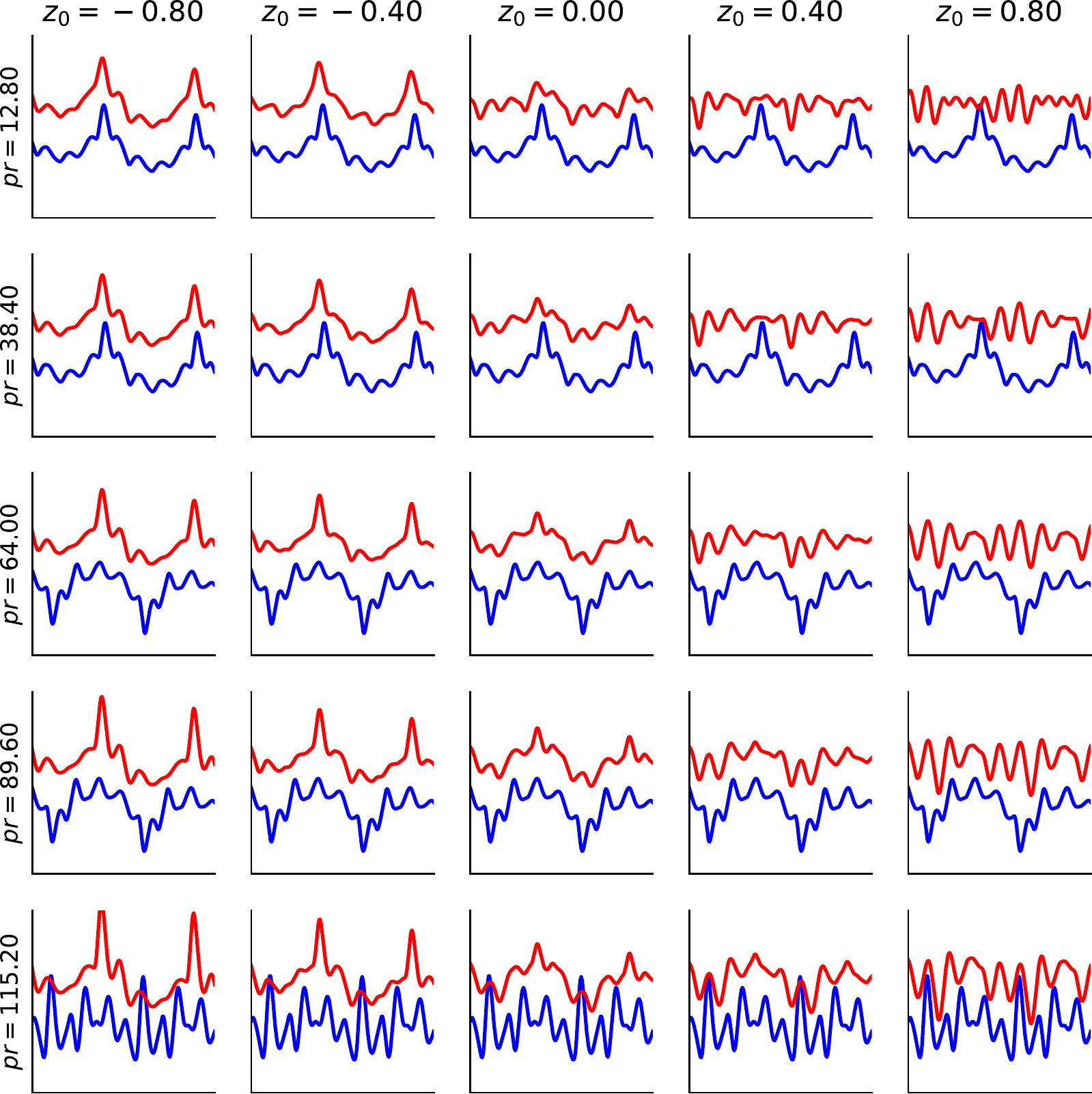} \hspace{0.2cm}
    (b) \includegraphics[height=7cm]{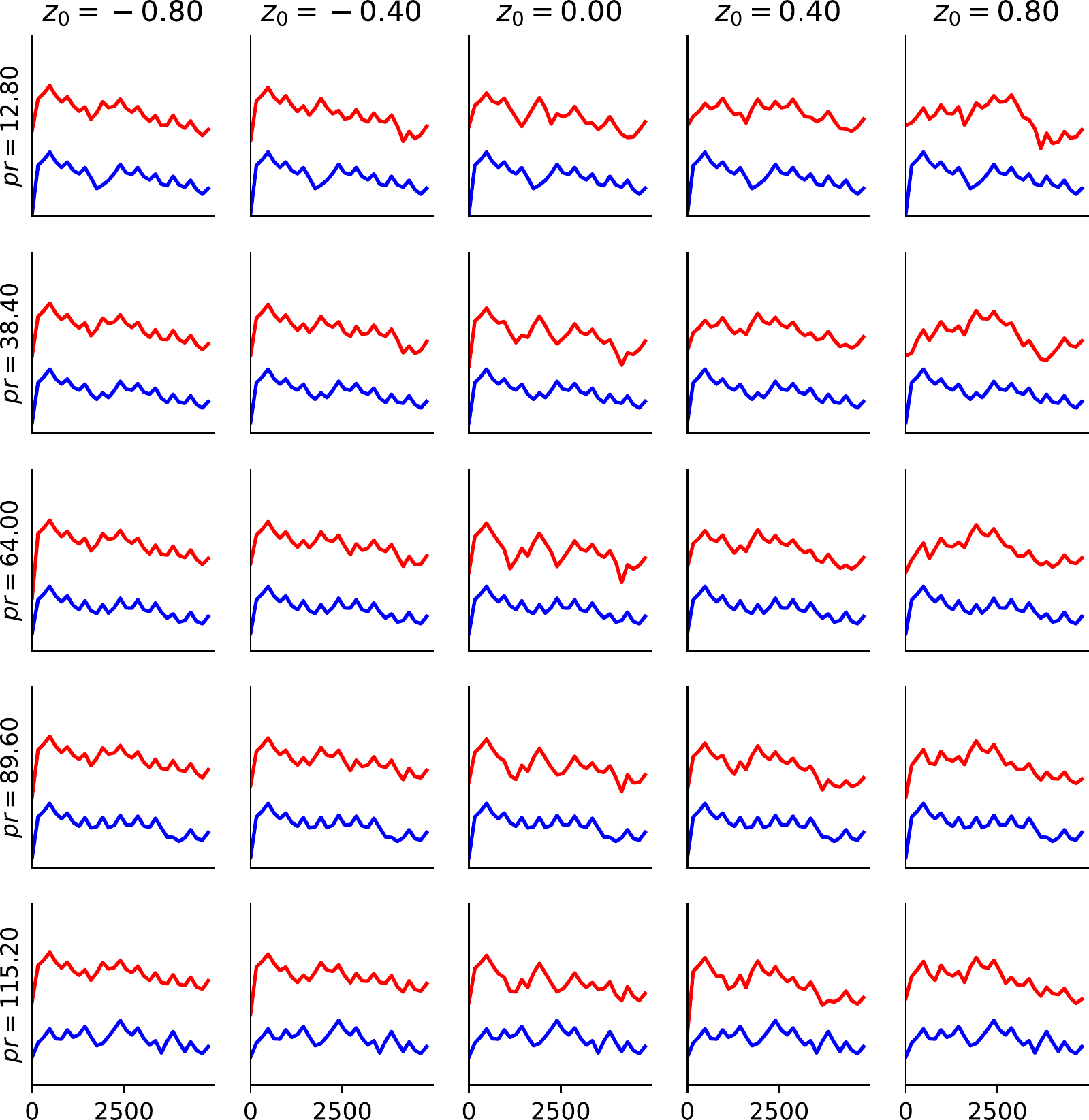}}
  \caption{ Output of $D1_Z2_Y$ as \emph{bow position} is set to 100,
    and \emph{bow pressure} and latent $z_0$ are changed. Top (red)
    Decoder output; bottom (blue) is the dataset. (a) Time domain; (b)
    Frequency domain. }
  \label{bowed-D1Z2Y-varyvz}
\end{figure}

\section{Results}\label{sec:results}

Figure~\ref{bowed-D1Z2Y-varypv} demonstrates the results of $D1_Z2_Y$.
Comparing the middle and bottom curves, we can see that while it has
some trouble with low values of \emph{bow pressure} and the extremes
of \emph{bow position}, the autoencoder is able to more or less encode
the distribution in our dataset.
The top curve (red) was generated by explicitly specifying the $y$ (conditional)
parameters instead of letting the autoencoder infer them, with $z_0=0$, and
demonstrates the output for parameter-driven reconstruction if $z_0$
is held constant.
Although not a perfect reproduction, particularly at extremes of the
\emph{bow position} range where there is more variance,
this demonstrates that the trained network is able to approximate
the data-parameter relationship present in the dataset.

\begin{figure}
  \centerline{
    (a)\includegraphics[width=0.45\textwidth]{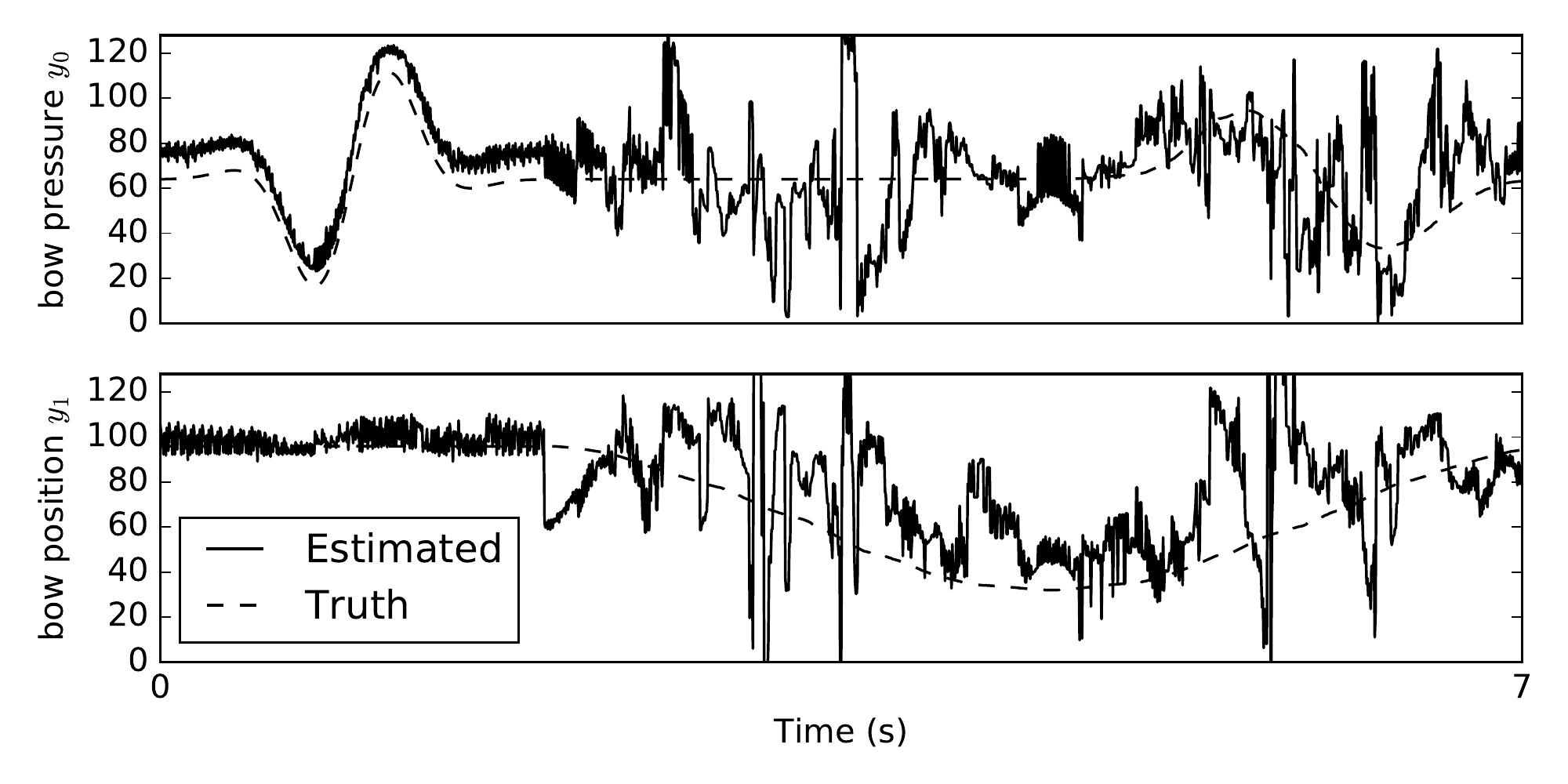}
    (b)\includegraphics[width=0.45\textwidth]{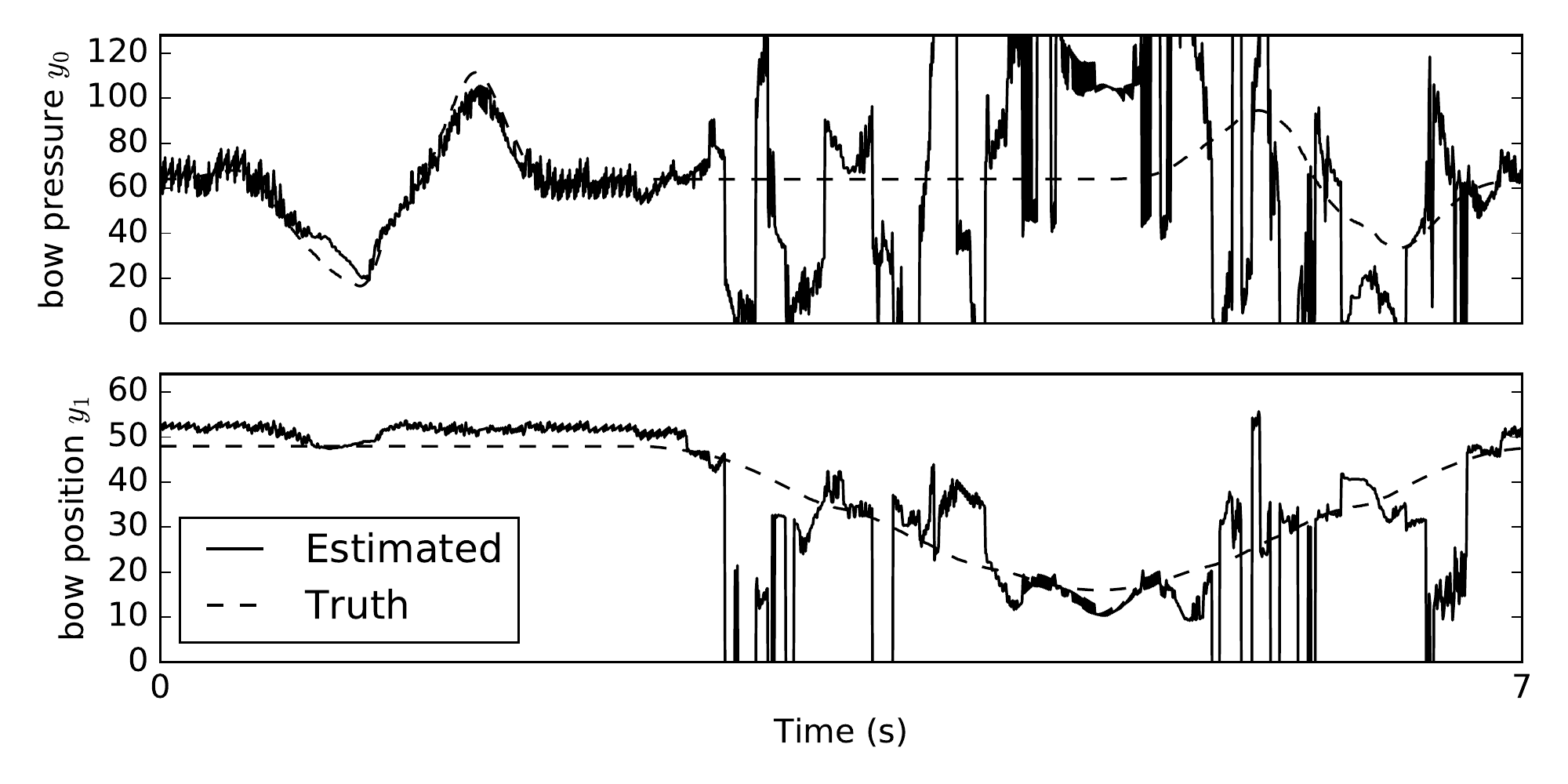}}
  \centerline{
    (c)\includegraphics[width=0.45\textwidth]{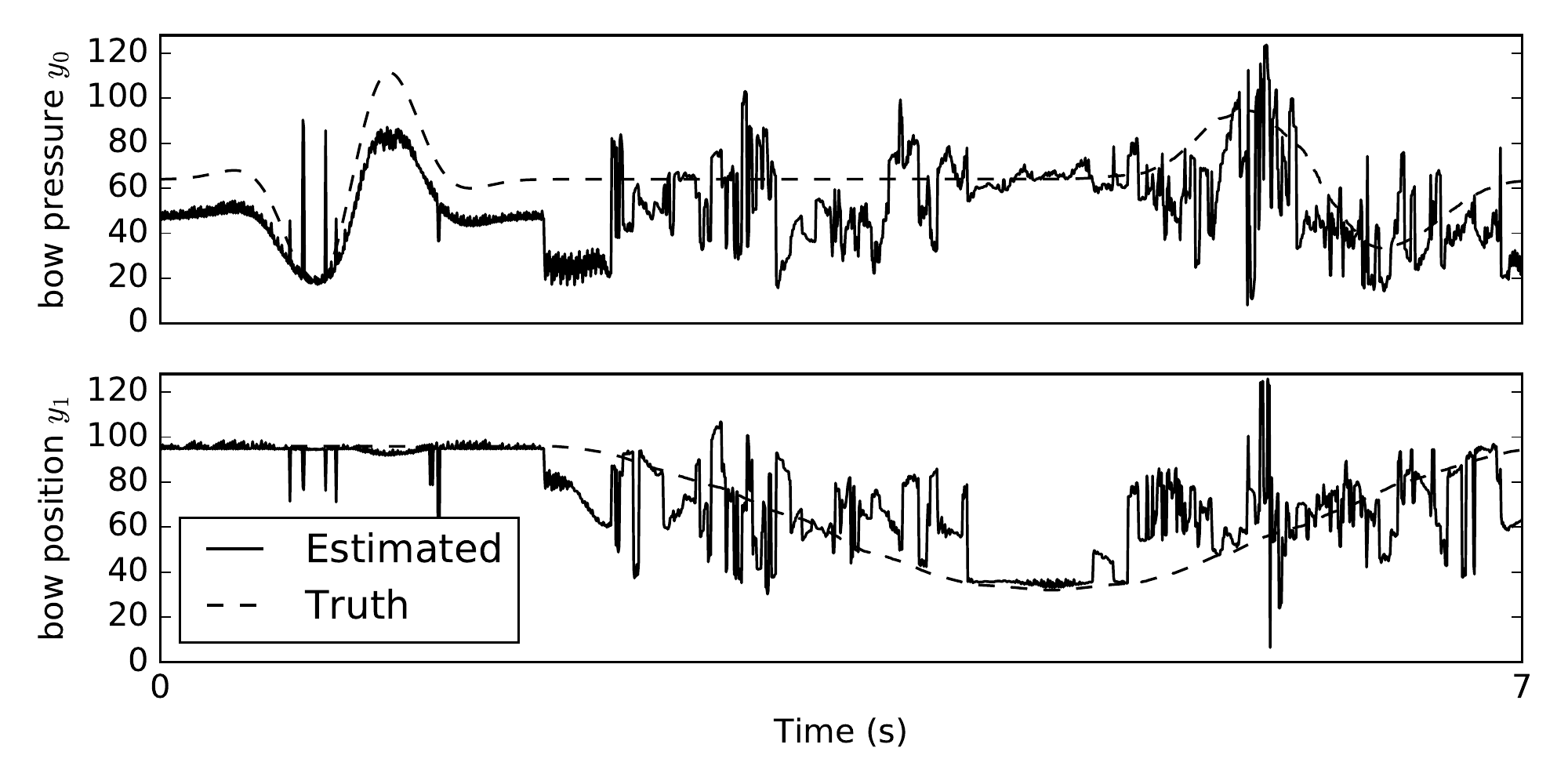}
    (d)\includegraphics[width=0.45\textwidth]{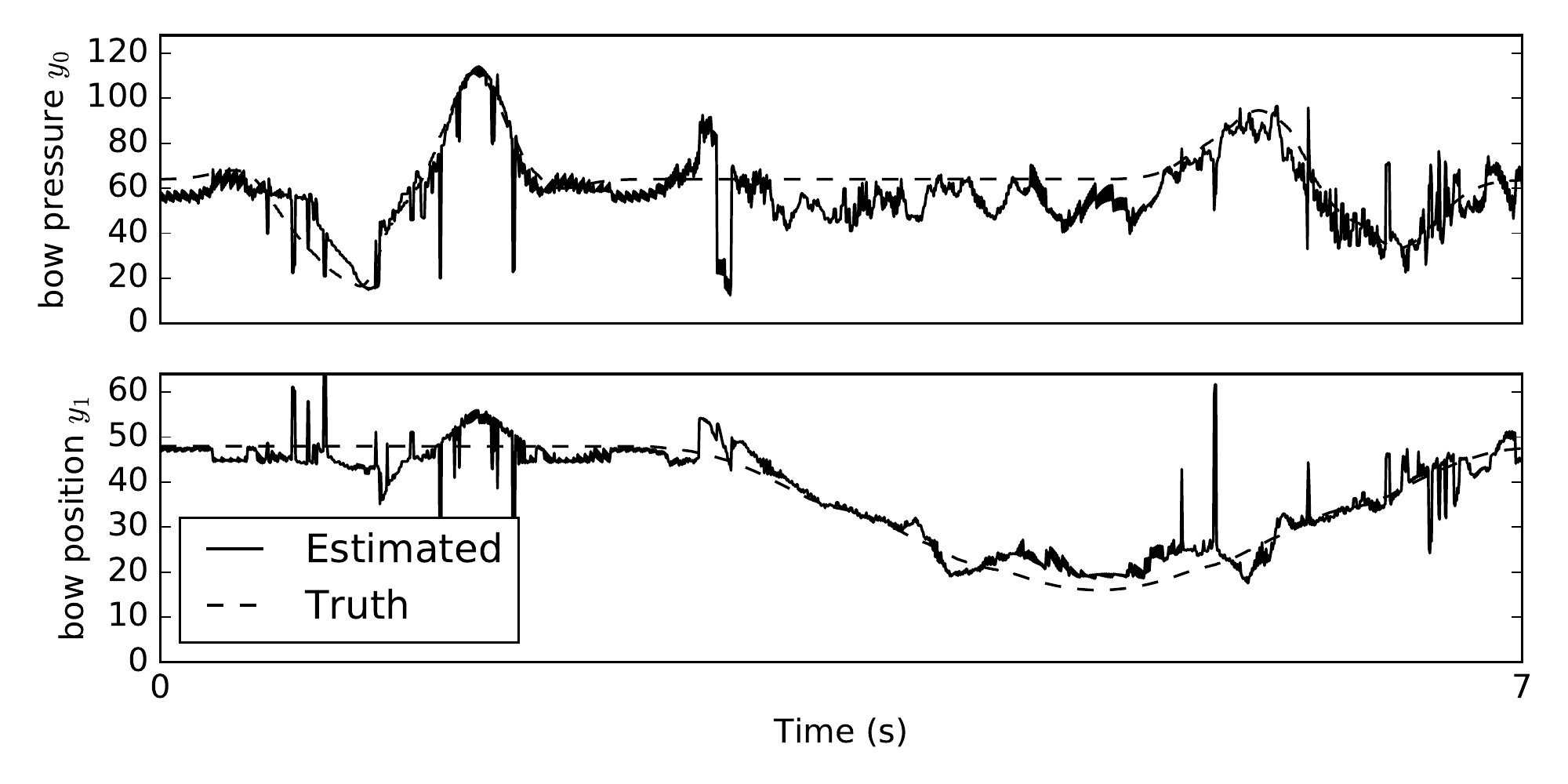}}
  \caption{ Parameter estimation performance of the $D1_Z2_Y$ network
    for (a)~\emph{bowed1} full dataset, RMS error=23.23;
    (b)~\emph{bowed1} half dataset, RMS error=33.45; (c)~\emph{bowed2}
    full dataset, RMS error=19.54; (d)~\emph{bowed2} half dataset, RMS
    error=8.61. }
  \label{bowed-known}
\end{figure}

The role of $z$ is now considered in Figure~\ref{bowed-D1Z2Y-varyvz},
by holding \emph{bow position} constant ($y_1=100$) and examining how
the signal changes with $z_0$.
One notices that for some values of $z_0$ the signal matches well, and
for others it varies from the target signal.
For example, we can see that in this case, high values of $z_0$ push
the signal towards two sharp peaks, while low values of $z_0$ tend
towards more oscillations; both $z_0=-0.8$ and $z_0=0.8$ resemble the
$pr=115.2$ condition, but in different aspects.
Meanwhile there is consistency with the ``stylistic'' influence of
$z_0$ on the signal for different values of bow pressure; for lack of
better words, in the time domain it changes from ``peaky'' to
``wiggly'' going from left to right.

Next, we look at the encoder (parameter estimator) performance, by
producing a \emph{new} signal from the STK synthesizer with a parameter
trajectory starting with smooth variation only in \emph{bow pressure} and
then smooth variation only in \emph{bow position}, and then in both
parameters.
Figure~\ref{bowed-known}(a) shows actually rather disappointing
performance in this respect, however it does clarify some information
not present in the previous analysis: the estimation is clearly better
for \emph{bow pressure}, but easily disturbed by changes in \emph{bow
  position}.
Nonetheless we see the tendency of the estimate in the right
direction, with rather a lot of flipping above and below the center.
Since varying the hyperparameters of our network did not solve this
problem, we hypothesized that this error could come from two sources:
(1) ambiguities in the dataset---indeed, if one examines the shape of
the signal as \emph{bow position} changes, one notices a symmetry
between values on either side of \emph{pos}=64, c.f.\ samples from
dataset in Figure~\ref{bowed-D1Z2Y-varypv}, blue line.
By consequence the inverse problem is underspecified, leading to
ambiguity in the parameter estimate.  (2) underrepresented variance in
the dataset; the new testing data varies continuously in the
parameters, but the dataset was constructed based on the per-parameter
steady state.

To investigate this, the network was trained on a ``half dataset'',
consisting only of samples of \emph{bowed1} where \emph{bow position}
$< 64$.
Furthermore, as mentioned, an extended dataset, \emph{bowed2}, was
constructed based on random parameter variations.
Results in Figure~\ref{bowed-known}(b)-(d) show that training on the
half-\emph{bowed1} dataset changed the character of errors, but did
not improve overall, however the extended \emph{bowed2} dataset gave
improved parameter estimation, and much improved in the \emph{bow
  position} $< 64$ case.
Thus it can be concluded that both sources contributed to parameter
estimation difficulties.

\begin{figure}[p]
  \centering
  \begin{tabular}{cc}
    (a) \emph{Unregularized} & (b) \emph{Adversarial regularization} \\
    \multicolumn{1}{l|}{\includegraphics[width=0.5\textwidth]{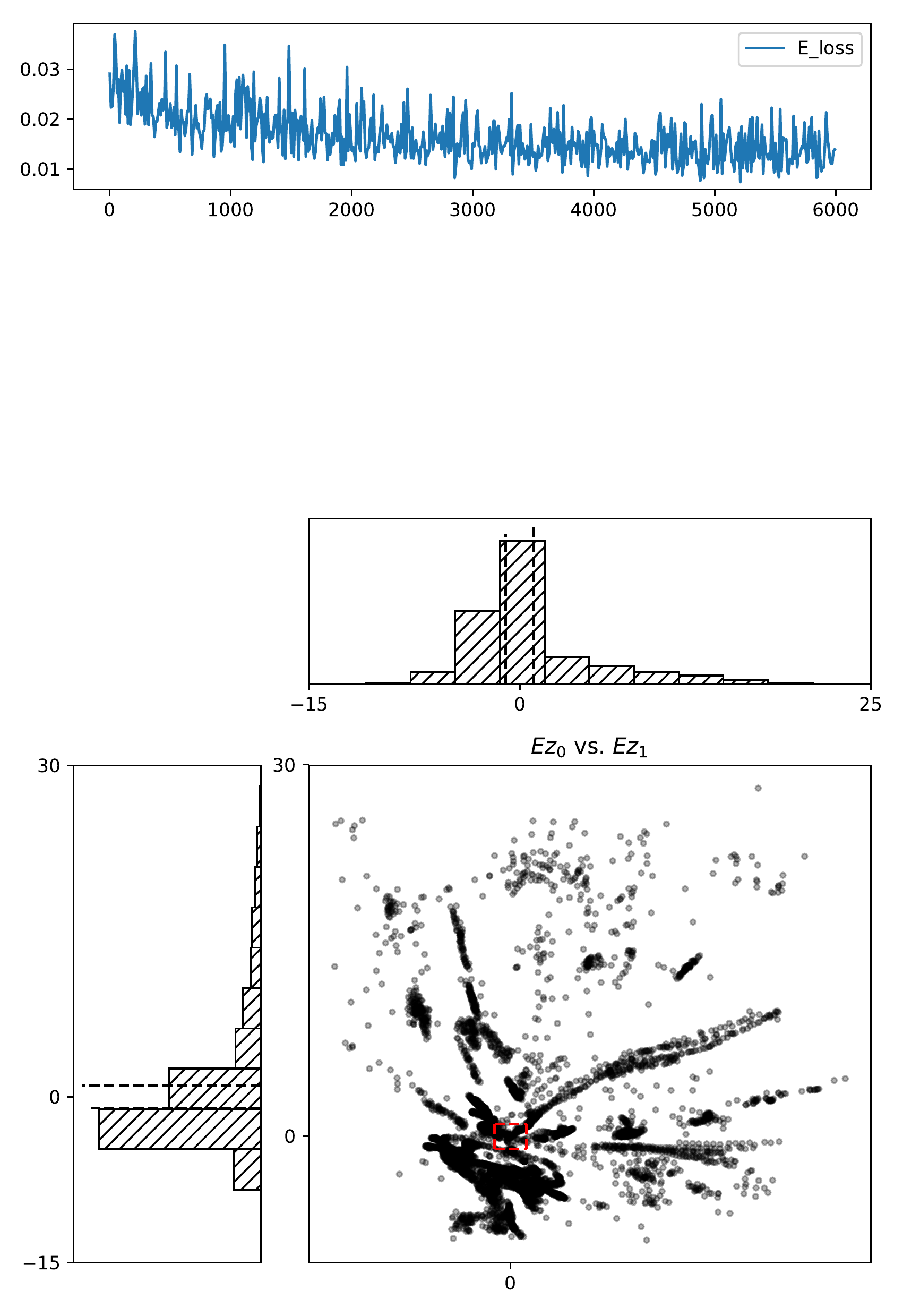}} &
    \includegraphics[width=0.5\textwidth]{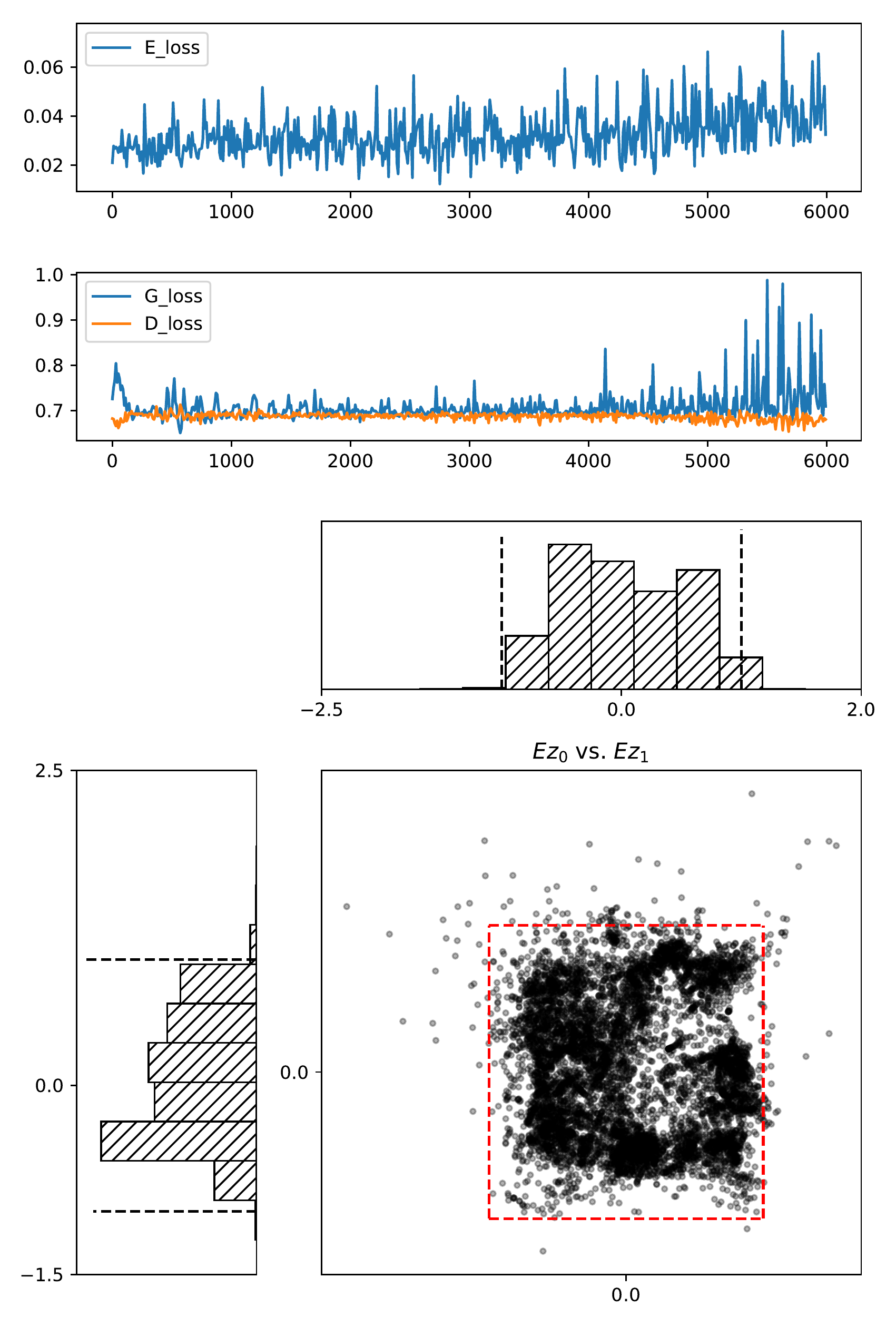} \\
    \multicolumn{1}{l|}{\includegraphics[width=0.5\textwidth]{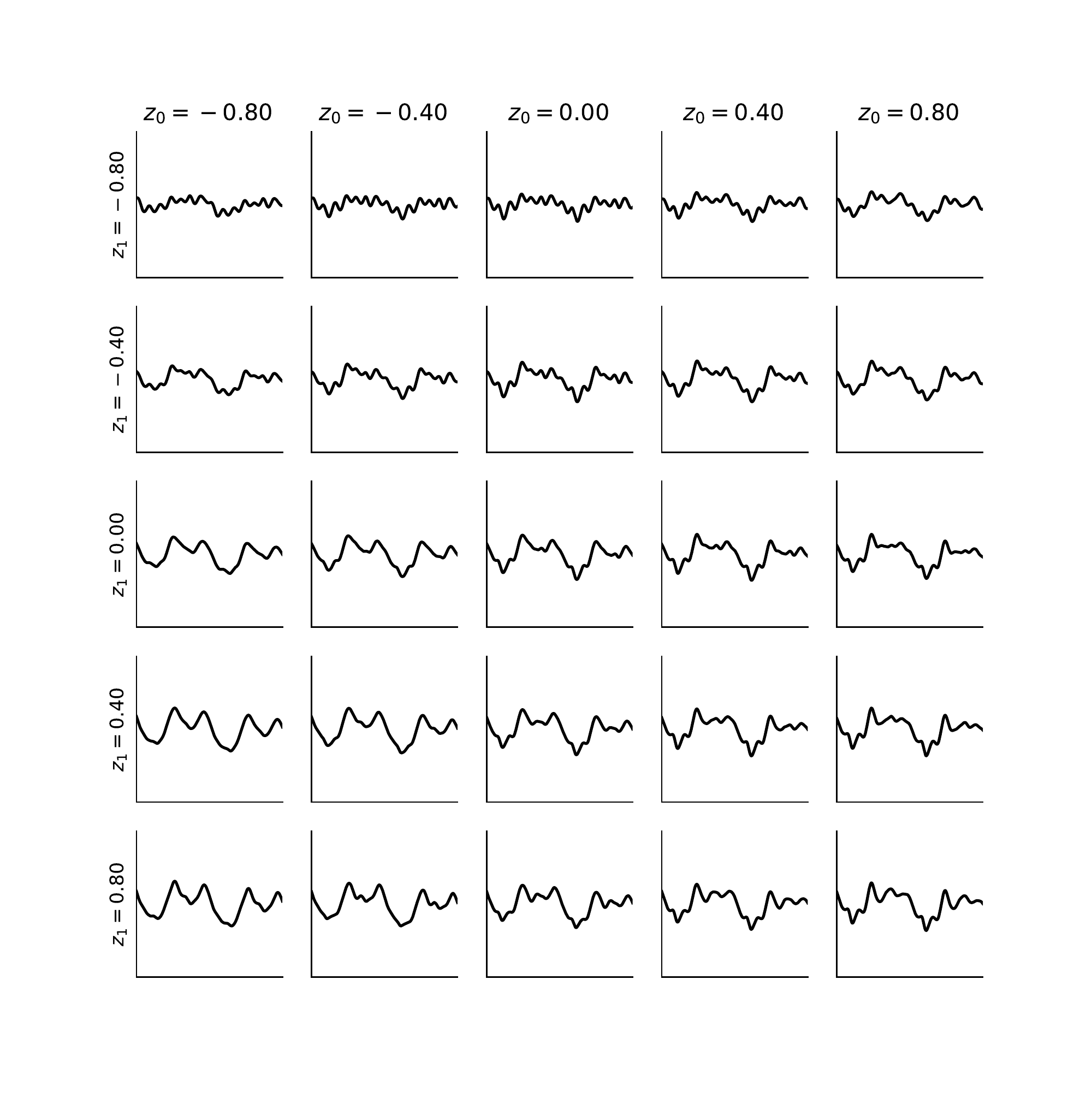}} &
    \includegraphics[width=0.5\textwidth]{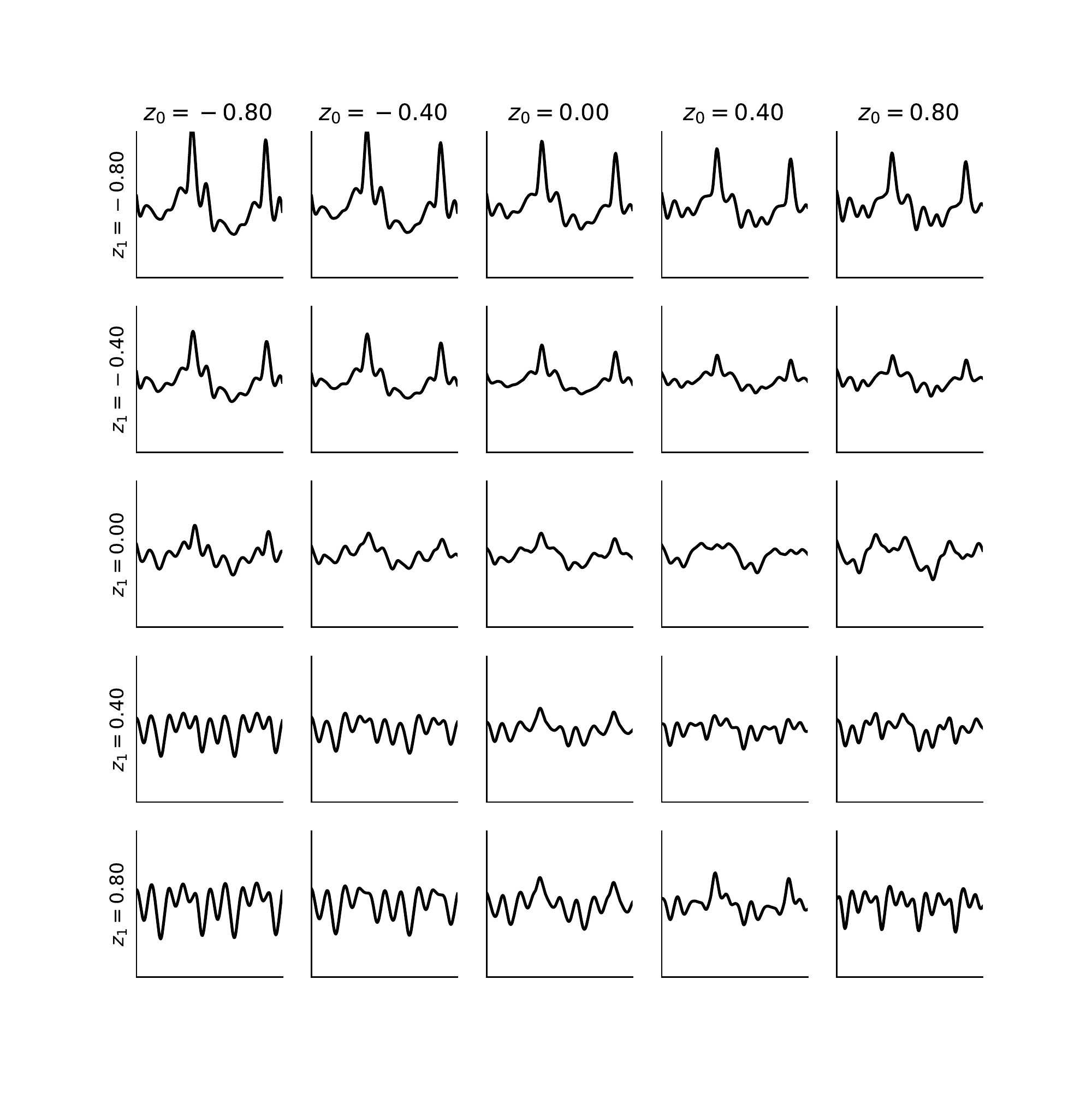} \\
  \end{tabular}
  \caption{ Distributions of latent parameters corresponding to a
    random sample of 3000 cycles from the dataset when trained (a)
    without regularization ($N2_Z0_Y$) and (b) with adversarial
    regularization ($D2_Z0_Y$).  The top rows indicate reconstruction
    error ($E_\textrm{loss}$, mean square error in normalized
    representation), and adversarial discrimination errors
    ($G_\textrm{loss}=L_G$, $D_\textrm{loss}=L_D$).  Regularization
    encourages the network to find a latent space such that the full
    variance of the dataset can be accessed uniformly from a
    restricted range of values, appropriate for interactive control.}
  \label{bowed-paramdist}
\end{figure}

Figure~\ref{bowed-paramdist} shows the resulting parameter space if
both parameters are left to be absorbed by the unsupervised latent
space.
The adversarial regularization regime can be seen in the generator and
discriminator losses of Fig.~\ref{bowed-paramdist}(b), which
encourages the autoencoder to make the distribution of these variables
similar to $\mathcal{U}(-1,1)$, i.e., a rectangle.
This facilitates user interaction with the generator, since
limited-range control knobs can be mapped to this rectangle, thus
having a strong chance to access the full range of variance present in
the dataset; conversely, the chance of synthesizing a sound that does
not correspond with the training data is minimized.
Without regularization, Fig.~\ref{bowed-paramdist}(a) ($N2_Z0_Y$), we
see some relationship between the two inferred variables $z_0$ and
$z_1$ (Fig.~\ref{bowed-paramdist})---although it appears more complex
than could be captured by a Pearson's correlation---while this is
completely gone for the regularized version ($D2_Z0_y$).
The spreading clusters are generated because without regularization,
the autoencoder attempts to maximally separate various aspects of the
variance in a reduced 2-dimensional space in order to decrease
uncertainty in reconstruction, which can be useful for data analysis
but does not produce a good interpolation space.
The regularization therefore encourages the parameter space to be interactively
``interesting,'' in the sense that the parameters represent orthogonal
(or at least, uncorrelated) axes within the distribution that cover a
defined domain (red square in Fig.~\ref{bowed-paramdist}) and tend
towards uniform coverage without ``holes''.

\begin{figure}[p]
  \centerline{\includegraphics[width=\textwidth,trim=1.3cm 0 2.6cm 1.2cm,clip]{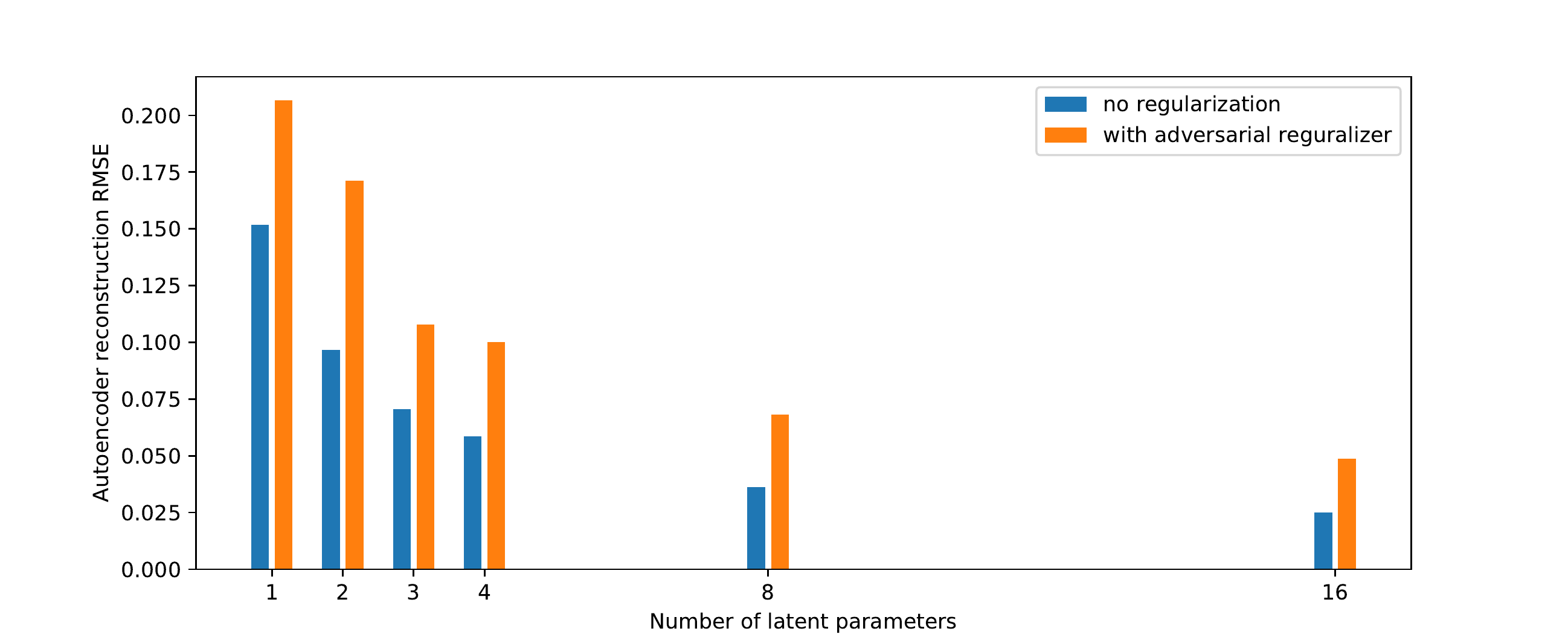}}
  \caption{Reconstruction mean squared error as a function of the
    number of latent parameters, with and without reguralization.}
  \label{latent-loss}

  \centering
  \begin{tabular}{cc}
    (a) \emph{Unregularized} & (b) \emph{Adversarial regularization} \\
    \multicolumn{1}{l|}{\includegraphics[width=0.48\textwidth,trim=10 0 10 270,clip]{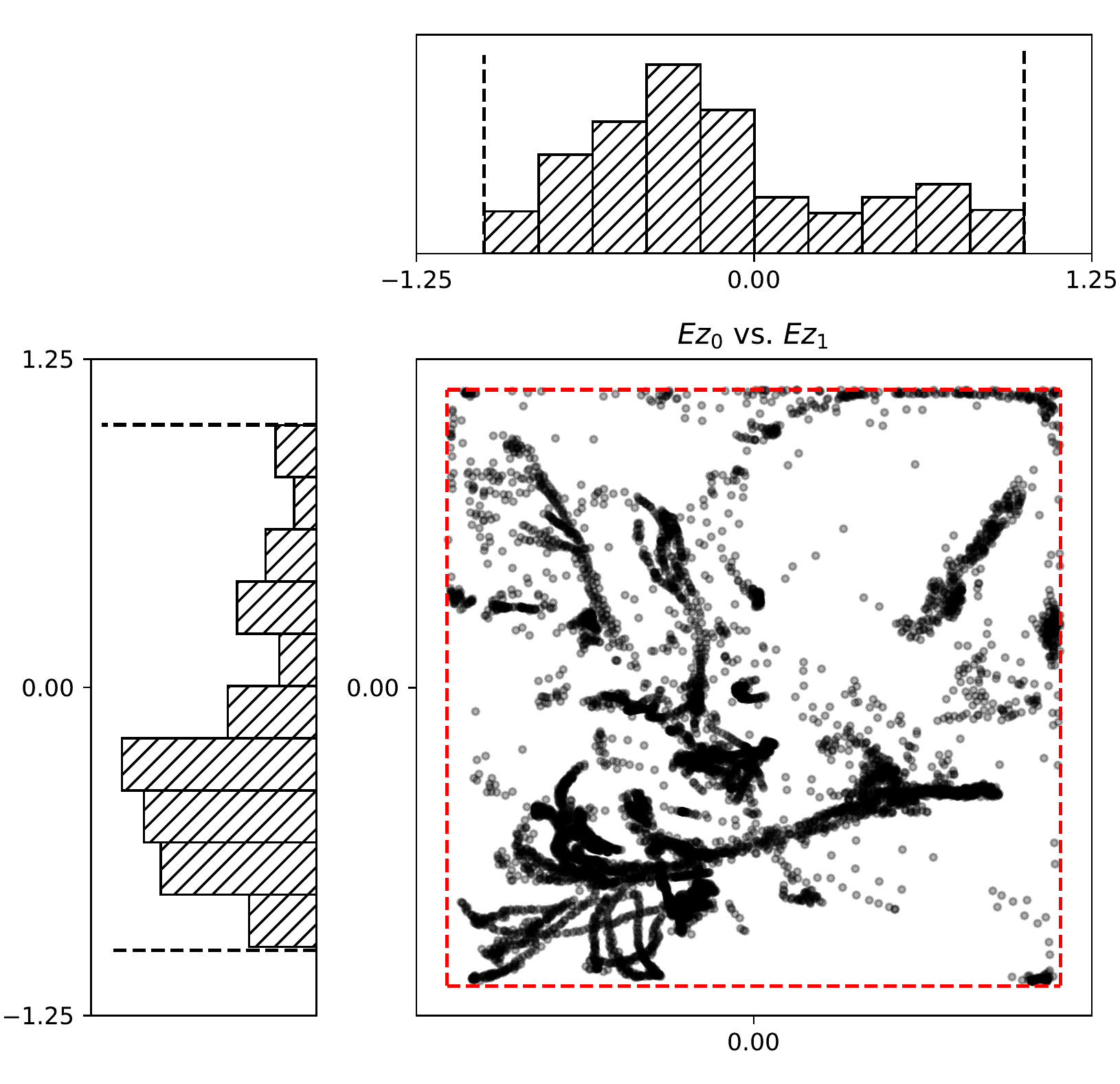}} &
    \includegraphics[width=0.48\textwidth,trim=10 0 10 270,clip]{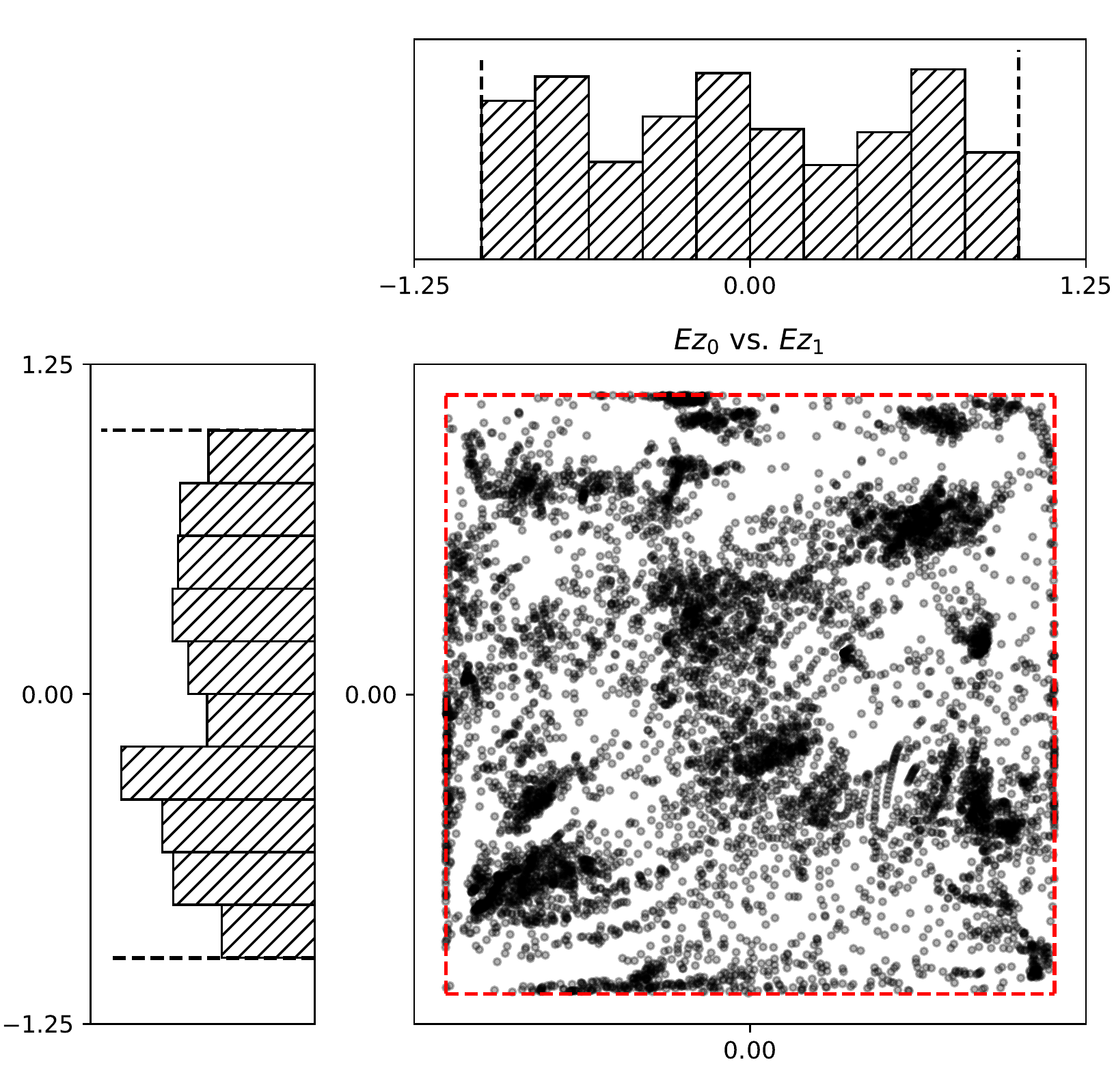}
  \end{tabular}
  \caption{ Results of same conditions as Fig.~\ref{bowed-paramdist},
    but the ReLU activation functions are replaced with the hyperbolic
    tangent in order to restrict the domain of $z$ instead of relying
    on the regularizer.  It can be seen that when the network
    architecture provides domain limiting, the regularization still
    provides a benefit of better approximating a uniform distribution,
    and further decorrelating the parameters.}
  \label{bowed-paramdist-tanh}

  \centerline{(a)\includegraphics[width=0.35\textwidth,trim=0 20 0 750,clip]{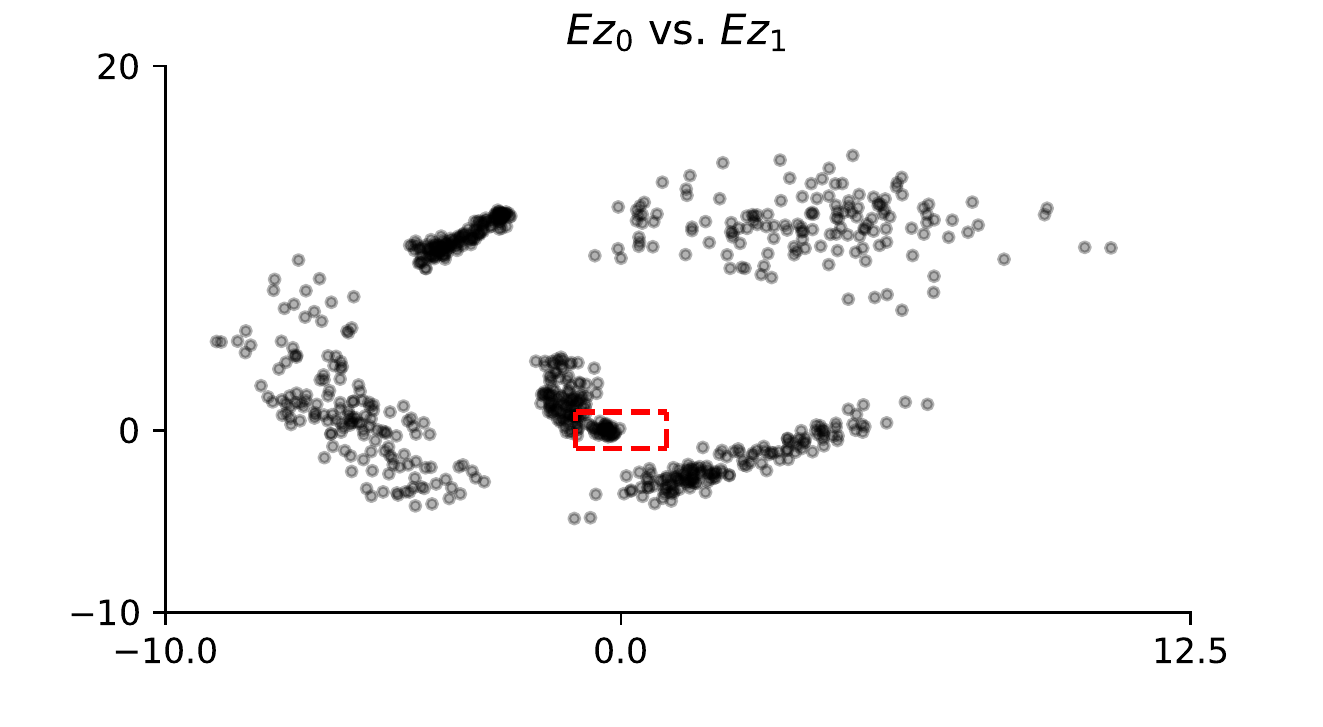}
  (b)\includegraphics[width=0.35\textwidth,trim=0 20 0 750,clip]{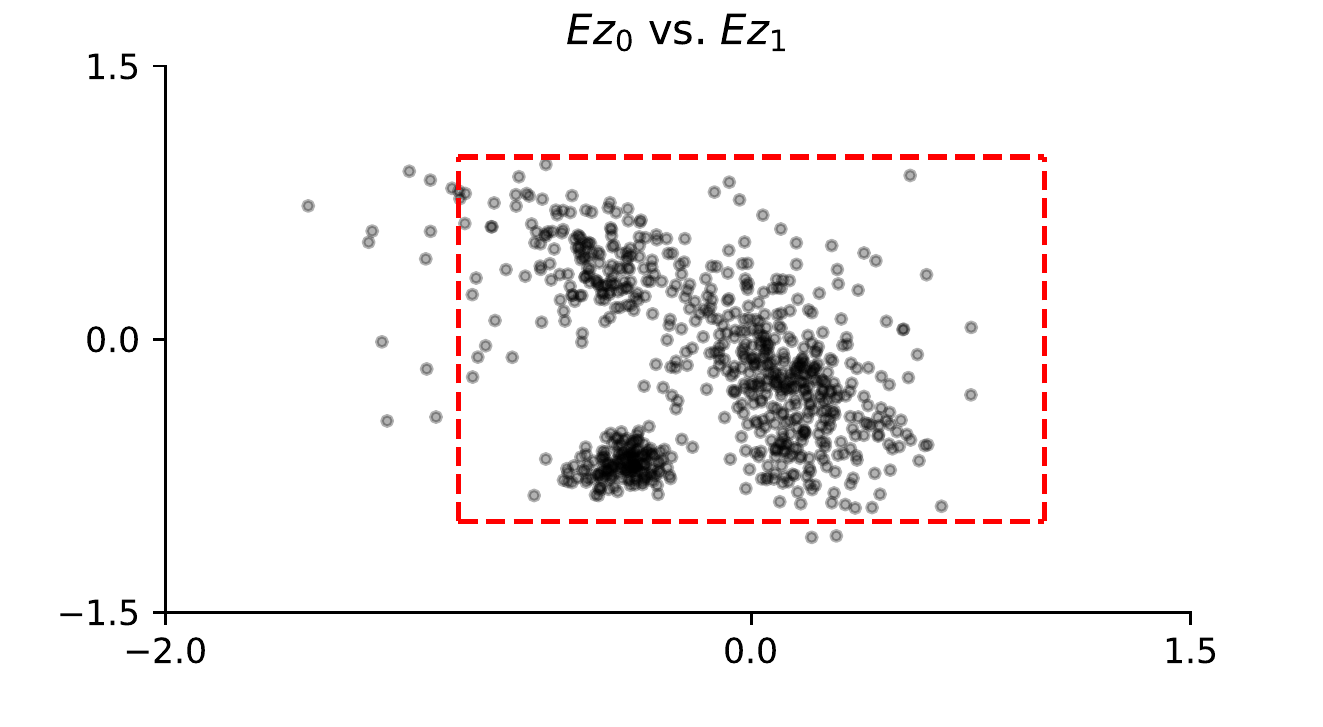}
  (c)\includegraphics[width=0.18\textwidth,trim=400 0 815 0,clip]{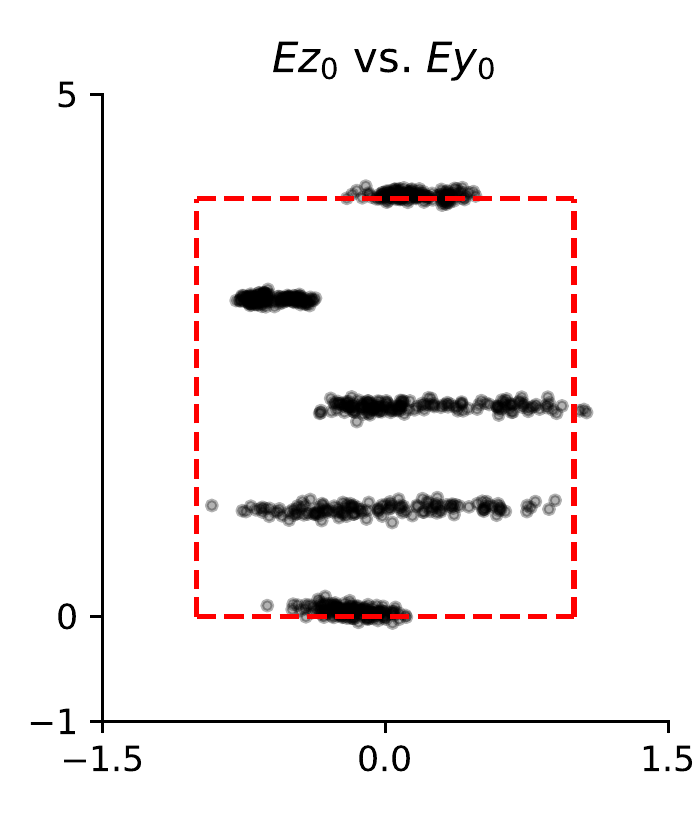}
  }
  \caption{Results on \emph{vowels} dataset: (a) using two latent
    parameters without regularization, the vowels are separated into
    clusters; (b) regularization encourages spreading to cover the
    uniform space within the desired boundaries, effectively smoothing
    out and blending the clusters; (c) replacing one latent parameter
    with conditioning on a vowel number 0 to 4.}
  \label{vowels-results}
\end{figure}

Of course, it is possible to restrict the domain without relying on
the regularizer, simply by defining the network architecture
accordingly.  For instance, if the non-linear units are changed for
the hyperbolic tangent, it is impossible for the network to generate
values outside the range $[-1,1]$.  In this sense the network
architecture itself can be understood as contributing to
regularization by enforcing hard constraints, rather than the soft
constraints of the cost function.  In Fig.~\ref{bowed-paramdist-tanh},
this \emph{tanh} architecture is demonstrated on the \emph{bowed1}
dataset, and it can be seen that the adversarial regularization
nonetheless is still useful for ensuring that the domain is used
effectively, i.e., despite some visual clusters still being apparent,
they are much more spread out, better approximating a uniform
distribution and, to a large degree, breaking up the piecewise
correlations between the parameters that can be seen by inspection
when regularization is not used.
In general, we found the ReLU approach more stable and better
at producing uniform coverage.

\begin{figure}
  \begin{tabular}{>{\centering}m{0.45\textwidth}>{\centering}m{0.45\textwidth}}
    {\includegraphics[width=0.45\textwidth]{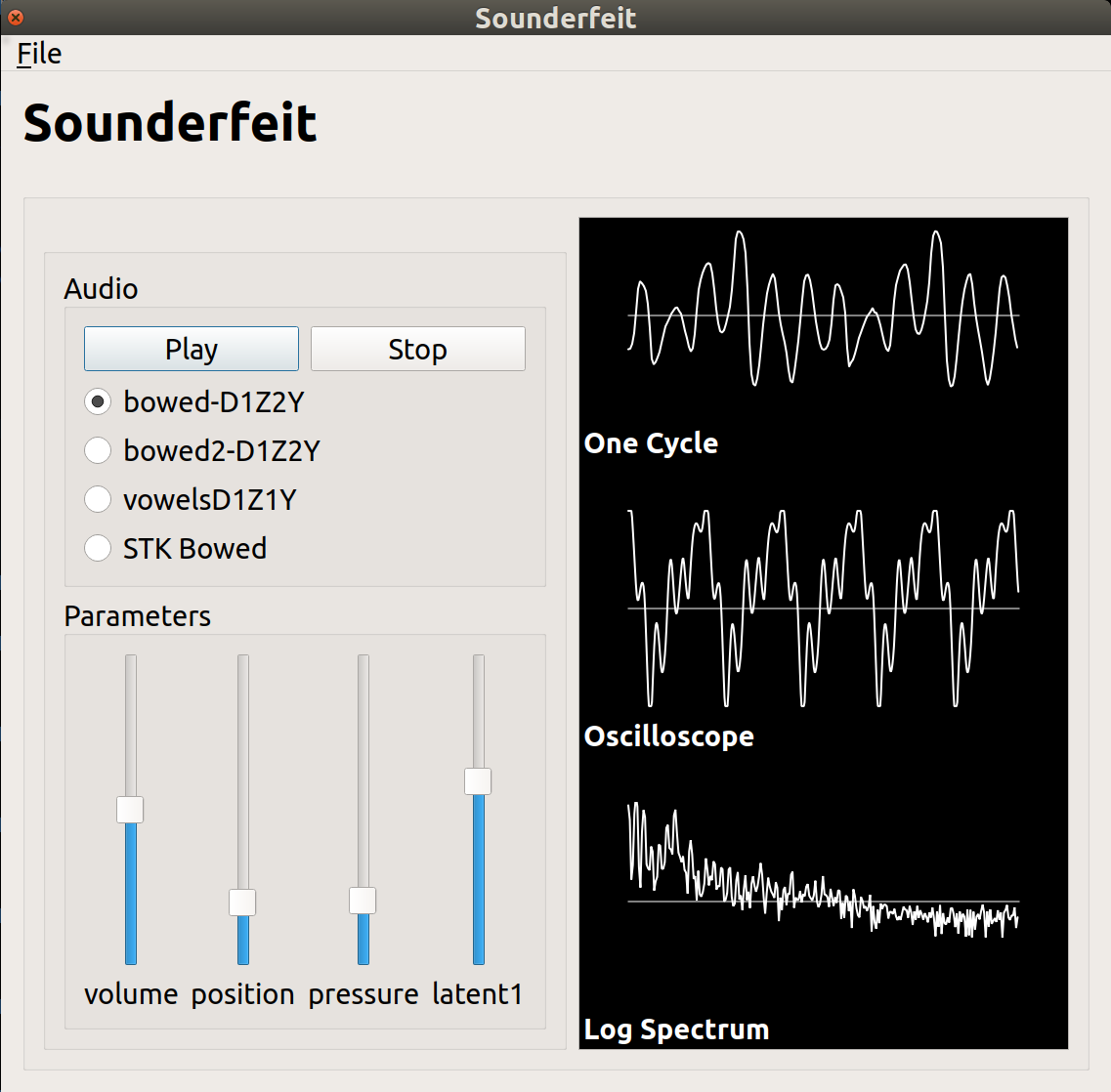}} &
    {\includegraphics[width=0.45\textwidth]{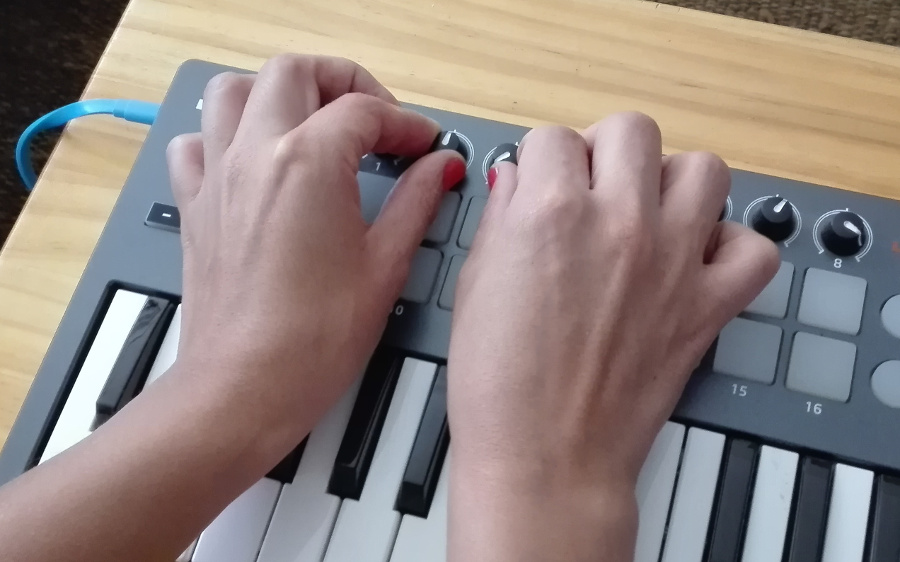}}
  \end{tabular}
  \caption{The Sounderfeit graphical user interface allows to
    interact in real time with the ANN-synthesized sound and compare
    it to the physical model that it was trained on.  Here, pressure
    and position are controlled by limited-range knobs on a MIDI
    keyboard (Novation).  }
  \label{sounderfeit}
\end{figure}

One will undoubtedly notice that the reconstruction error as reported
in Fig.~\ref{bowed-paramdist} does suffer due to the reguralization.
Indeed this is an expected outcome since the regularization applies
extra requirements such that the training will sacrifice one criteria
to improve another.  Additionally, the error will depend greatly on
how much variance is present in the data vs.\ how much ``room'' it
needs to express it---in this sense, we would expect accuracy to
increase as latent dimensions are added.  Fig.~\ref{latent-loss} gives
an idea of how reconstruction error changes as we do so.

\begin{figure}[t]
  \centerline{
    \includegraphics[width=\textwidth]{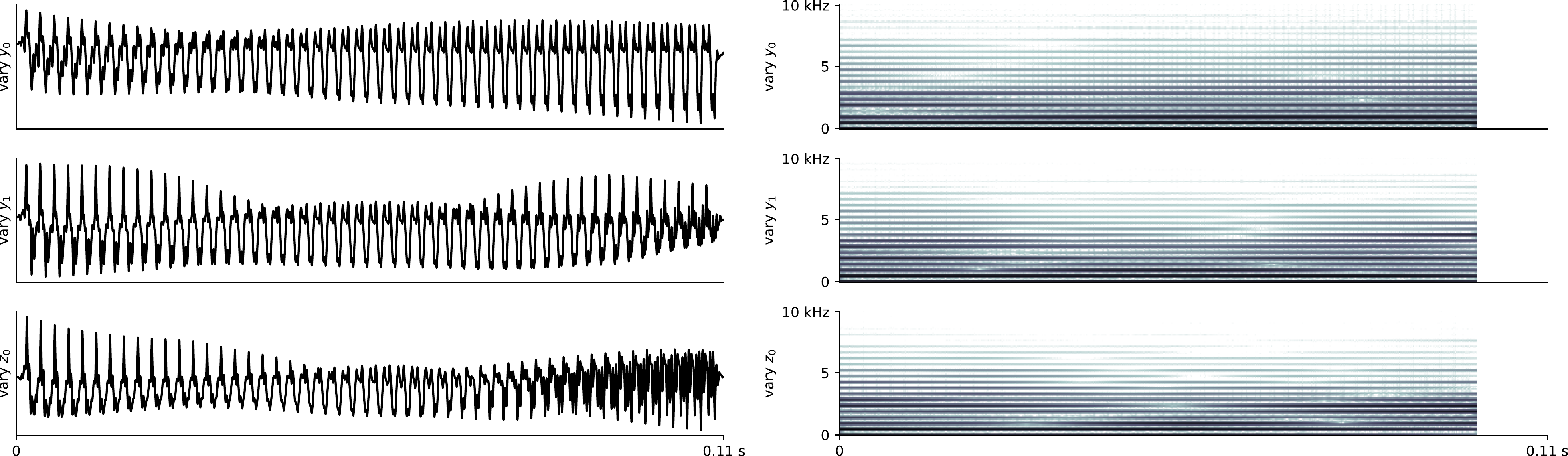}}
  \caption{ Overlap-add output of $D1_Z2_Y$, varying each parameter
    over a short interval. }
  \label{bowed-sound}
\end{figure}

We found that with this small decoder network of 20300 weights and 300
biases, an overlap-add synthesis could be performed in real time on a
laptop computer (10 seconds took 8.5 seconds to generate and was much
faster when re-implemented in C++), and
we can thus present a real-time, interactive data-driven wavetable
synthesizer, which we call Sounderfeit, see Figure~\ref{sounderfeit},
with a number of adjustable parameters.\footnote{Sounderfeit source
  code can be found on its project page at
  \url{https://gitlab.com/sinclairs/sounderfeit}}
The output of the overlap-add process is visualised in
Figure~\ref{bowed-sound}.

Lastly, in order to verify this method on another dataset, a similar
network was trained on the \emph{vowels} dataset, adjusted to have the
right size of input layer of 800, see Figure~\ref{vowels-results}.
The inferred space reflects the condition number (discrete, here), but
the remaining parameter uniformly covers the range $[-1,1]$. In this case the
extra variance beyond the conditional parameter consists only of small
tonal changes in the recorded voice as well as microphone noise, and
thus there is much more variance between vowels than within. It is
apparent from Fig.~\ref{vowels-results}(c) that there is no
``leakage'' of these extra sources of variance to the vowel label
parameter axis, as the vowel number adequately identifies the
cluster. The clusters can therefore be nicely mapped to a desired
space automatically by the conditioning and regularization.

\section{Conclusions}

These experiments showed some modest success in copying the
parameter-data relationship of a physical modeling synthesizer
and fitting them into a desired configuration.
Like many machine learning approaches, the quality of results depends
strongly on the hyperparameters used: network size and architecture,
learning rates, regularization weights, etc., and these must be
adapted to the dataset.
Shown here are results from the best parameters found after some
combination of automatic and manual optimisation on this specific
dataset, which we use to demonstrate some principles of the design,
however it should be noted that actual results varied sometimes
unexpectedly with small changes to these parameters.
This hyperparameter optimization is non-trivial, especially when it
comes to audio where mean squared error may not reveal much about the
perceptual quality of the results, and so a lot of trial and error is
the game.
Thus, a truly ``universal'', turn-key synthesizer copier would require
future work on measuring a combined hypercost that balances well the
desire for good reproduction with good parameter estimation quality,
and well-distributed latent parameters.  Such work could go beyond
mean squared error to involve perceptual models of sound perception.
For example, recent work in speech synthesis has shown a significant
improvement in perceived quality when the model was conditioned on
mel frequency spectrograms \cite{shen2017}.

Some practical notes: (1) We found that getting the adversarial method
to properly regularize the latent variables in the presence of
conditional variables is somewhat tricky; the batch size and relative
learning rates played a lot in balancing the generator and
discriminator performances.  New research in adversarial methods is a
current area of investigation in the ML community and many new
techniques could apply here; moreover comparison with variational
methods is needed---we note however that variational autoencoders are
typically regularized to fit a Gaussian normal distribution, whereas
an advantage of the adversarial approach is to fit any example-based
distribution, which we took advantage of to fit the rectangle
accessible by a pair of knobs.
(2) We found the parameter estimation extremely sensitive to phase
alignment; we tried randomizing phase of examples during training,
which gave better parameter estimates, but this was quite damaging to
the autoencoder performance.  In general oversensitivity to global phase is a
problem with this method, a downside to the time domain
representation; more work on dealing with phase as a latent parameter
is necessary.

Nevertheless we have attempted to outline some potential for use of
autoencoders and their latent spaces for audio analysis and synthesis
based on a specific signal source.
Only a very simple fully-connected single-layer architecture was
used, and thus improvements should be explored,
in particular the addition of convolutional layers.
The advantages of time- and frequency-domain representations as
learning targets should be characterised.
More important than the quality of these specific results, we wish to
point out the modular approach that autoencoders enable in modeling
oscillator periods of known and unknown parameters, and that, in
contrast to larger datasets covering many instruments
\cite{engel2017neural}, interesting insights and useful performance
systems can be generated even from small data.

One might ask what a blackbox model brings to the table in the
presence of an existing, semantically-rich physical model.
Indeed, in this work a digital synthesizer was used as an easy way to
gain access to a fairly complicated but clean signal with a small
number of parameters.
In principle this method could be used on much richer, real instrument
recordings.
To demonstrate this point we have trained it, as shown, on a very
small vocal recording (3 seconds per vowel) and produced a working
vowel synthesizer with separate knobs for vowel number and ``noise'',
e.g.\ microphone noise and vocal variance, however more complex
experiments are needed in this vein.  A challenge in operating on real
data was cutting and aligning oscillation cycles correctly, which was
non-trivial and prohibits easier experimentation on arbitrary data
streams.

Simultaneous estimation and generation with the same network may be
unnecessary.
In fact conditioning variables could be made external inputs instead
of inferred from the input data, and the decoder could be used
separately to only learn the parameters in a more typical regression
configuration.
However, one of the long-term goals of performing automatic inference
is to play somewhat with the latent and parameter space, such as using
it for what is known in the audio community as cross-synthesis, or in
the machine learning community as ``style transfer'', i.e., swapping
the bottom and top halves of two such autoencoder networks, allowing
to drive a synthesizer by both conditioned and latent parameters
estimated on an incoming signal.
One can imagine, for example, playing the violin and having the bow
pressure control the brightness of a wind instrument sound, while
subtle aspects of the gesture are left to latent space to control
more subtle parameters of the sound.  To achieve this a much less
noisy inference result would be necessary, and is of course predicated
on the idea that the parameter to data function is invertible, which,
as seen in our failure to map the complete \emph{bow position} domain,
is not necessarily a given.  Another reason for doing simultaneous
estimation and generation left for future work is to investigate
whether a tied-weights approach might improve both goals by
integrating mutual sources of information on either side of the
equation.

\begingroup
\parindent 0pt
\parskip 2ex
\def\enotesize{\footnotesize}
\def\notesname{Note}
\theendnotes
\endgroup


\providecommand{\abntreprintinfo}[1]{%
  \citeonline{#1}}
\setlength{\labelsep}{0pt}

\vspace*{\fill}
\noindent\rule\textwidth{2\arrayrulewidth}
\noindent{\scriptsize
  \textbf{Stephen Sinclair} – In 2012 completed a PhD at McGill
  University in Montreal. His topic was audio-haptic interaction with
  musical acoustic models. He spent 3 years as a post-doctoral
  researcher in the ISIR laboratory of UPMC, Paris, working on new
  haptic interaction methods. Currently is a research engineer at
  Inria Chile, principally working to improve the Siconos non-smooth
  dynamical system simulation engine.\vspace{-0.15cm}\\
  \rule\textwidth{2\arrayrulewidth}\par}\vspace{1.3cm}

\pagestyle{last}

\end{document}